\documentstyle[twoside,fleqn,espcrc2]{article}
\input epsf
 
\def\mapright#1{\!\!\!\smash{
\mathop{\longrightarrow}\limits^{#1}}\!\!\!}

\newlength{\extraspace}  
\newlength{\extraspaces}
\newcommand{\be}{\begin{equation}
\addtolength{\abovedisplayskip}{\extraspaces}
\addtolength{\belowdisplayskip}{\extraspaces}
\addtolength{\abovedisplayshortskip}{\extraspace}
\addtolength{\belowdisplayshortskip}{\extraspace}}
\newcommand{\ee}{\end{equation}}
\newcommand{\ba}{\begin{eqnarray}
\addtolength{\abovedisplayskip}{\extraspaces}
\addtolength{\belowdisplayskip}{\extraspaces}
\addtolength{\abovedisplayshortskip}{\extraspace}
\addtolength{\belowdisplayshortskip}{\extraspace}}

\newcommand{\ea}{\end{eqnarray}}
\newcommand{\diag}{{\rm diag}}
\newcommand{\tr}{{\rm tr}}

\newcounter{fignum}
\newcommand{\figuurnum}{\arabic{fignum}}

\newcommand{\figuurplus}[3]{
\addtocounter{fignum}{1}
\addcontentsline{lof}{figure}{\protect
\numberline{\arabic{section}.\arabic{fignum}}{#3}}
\hspace{-3mm}{\it fig.}\ \figuurnum.
\begin{figure}[t]\begin{center}
\leavevmode\hbox{\epsfxsize=#2 \epsffile{#1.eps}}\\[5mm]
\parbox{6.5cm}{\small \bf Fig.\ \figuurnum: \it #3} 
\vspace{-9mm}
\end{center} \end{figure}}
\newcommand{\cH}{{\cal H}}
\newcommand{\nonu}{\nonumber\\[2mm]}
\newcommand{\is}{& \!\! = \!\! &}
\newcommand{\twomatrix}[4]{{\left(\begin{array}{cc}#1 & #2\\
#3 & #4 \end{array}\right)}}
\newcommand{\twomatrixd}[4]{{\left(\begin{array}{cc}
\displaystyle #1 & \displaystyle #2\\[2mm]
\displaystyle  #3  & \displaystyle #4 \end{array}\right)}}
\newcommand{\ie}{{\it i.e.\ }}
\newcommand{\hf}{{\textstyle{1\over 2}}}
\newcommand{\half}{{\textstyle{1\over 2}}}

\newcommand{\adot}{{\dot{a}}}

\newcommand{\Z}{{\bf Z}}

\newcommand{\delbar}{\overline{\partial}}

\newcommand{\Lbar}{{\overline{L}}}

\def\a{\alpha} 
\def\b{\beta}
\def\th{\theta}
\newcommand{\del}{\partial}
\def\G{\Gamma}

\newcommand{\bps}{_{{}_{\rm BPS}}}

\newcommand{\wW}{{{\small  \cal W}}}
\newcommand{\kK}{{{\small \cal K}}}
\newcommand{\lL}{{\mbox{\small  \cal L}}}
\newcommand{\gG}{{\mbox{\small \cal F}}}
\newcommand{\sS}{{\eta}}

\renewcommand{\d}{{\partial}}

\newcommand{\Sym}{S}

\newcommand{\xX}{Y}

\newcommand{\R}{{\bf R}}

\newcommand{\ww}{{\bf 1}}

\newcommand{\ra}{\rightarrow}

\newcommand{\qf}{q_{{}_{ 5}}}
\newcommand{\qq}{{\bf q}}
\newcommand{\A}{{{}^{ A}}}
\newcommand{\B}{{{}^{ B}}}
\newcommand{\Aa}{{{}_{\strut \! A}}}
\newcommand{\Bb}{{{}_{\strut \! B}}}
\newcommand{\NS}{N\! S}
\newcommand{\KK}{K\! K}

\title{Notes on Matrix and Micro Strings\thanks{Based on lectures
given by H.V. at the APCTP Winter School held in Sokcho, Korea, 
February 1997, and joint lectures at Cargese Summer School June 1997, 
as well as on  talks given by H.V. at SUSY '97, Philadelphia, May 27-31, 
and by R.D. and E.V. at  STRINGS '97, Amsterdam, June 16-21, 1997.}}

\author{Robbert Dijkgraaf \address{Department of Mathematics\\[-.5mm]
University of Amsterdam, 1018 TV Amsterdam}, 
E. Verlinde
\address{TH-Division, CERN, CH-1211 Geneva 23, \\
Institute for Theoretical Physics\\[-.5mm]
University of Utrecht, 3508 TA Utrecht}, 
and H. Verlinde \address{Institute for Theoretical Physics\\[-.5mm]
University of Amsterdam, 1018 XE Amsterdam}}
\begin{document}

\begin{abstract}
We review some recent developments in the study of M-theory 
compactifications via Matrix theory. In particular we highlight 
the appearance of IIA strings and their interactions, and explain 
the unifying role of the 
M-theory five-brane for describing the spectrum of the $T^5$ 
compactification and its duality symmetries. The 5+1-dimensional 
micro-string theory that lives on the fivebrane world-volume 
takes a central place in this presentation.

\end{abstract}

\maketitle

\section{Introduction}

In spite of considerable recent progress, M-theory is as yet a theory
without a fundamental formulation. It is usually specified 
by means of a number of fundamental properties that we know it should
posses. Among the most important of these characteristics is that the 
degrees of freedom and interactions of M-theory should provide a 
(mathematically consistent) representation of the maximally 
extended supersymmetry algebra of eleven dimensional supergravity
\cite{democracy}
\ba
\label{1}
\{ Q_{\a}, Q_{\b} \} \is (\G^m)_{\a \b} P_m +
(\G^{mn})_{\a \b}Z^{(2)}_{mn} \nonu
&& \ \ + \; 
(\G ^{m_1\dots m_5})_{\a \b}Z^{(5)}_{m_1\dots m_5}\, .
\ea
Here $P_m$ denotes the eleven dimensional momentum, and $Z^{(2)}$
and $Z^{(5)}$ represent the two- and five-index central charges
corresponding to the two types of extended objects present in 
M-theory, respectively the membrane and the fivebrane. Upon 
compactification, these extended objects give rise to a rich spectrum 
of particles, with many quantized charges corresponding to the 
various possible wrapping numbers and internal Kaluza-Klein momenta.

An alternative but presumably equivalent characterization of M-theory is 
that via compactification on a circle $S^1$ it becomes equivalent to ten 
dimensional type IIA string theory \cite{witten}. The particles with 
non-zero KK-momentum 
along the $S^1$ are in this correspondence identified with the D-particles 
of the IIA model, while the string coupling constant $g_s$ 
emerges as the radius $R$ of the $S^1$ via
\be
\label{scales}
\ell_s g_s= R,\qquad \ell_s^2= {\ell_p^3/ R}.
\ee
Here $\ell_s$ denotes the string scale and $\ell_p$ is the 11-dimensional 
Planck scale. Also upon further compactification, all
M-theoretic charges now have well understood interpretations in terms
of perturbative and non-perturbative string theoretic configurations, 
such as wrapped strings, D-branes \cite{polch} or solitonic five-branes 
\cite{fivebrane}. The precise and well-defined perturbative rules for 
determining the interactions among strings and D-branes \cite{polch}
\cite{bound} \cite{ulf} \cite{probe} so far provides the best basis for concrete 
quantitative studies of M-theory dynamics. String theory still forms 
the solid foundation on which the new structure of M-theory needs to be 
built.

From both the above starting points, a convincing amount of evidence has 
been collected to support the conjecture that toroidal M-theory 
compactifications exhibit a large group of discrete duality 
symmetries, that interchange the various internal charges 
\cite{hulltownsend}. While some of these dualities
are most manifest from the eleven dimensional perspective, others become 
visible only in the string formulation. In combination, these so-called 
U-duality symmetries put very strong restrictions on the possible 
non-perturbative properties of M-theory, and in this way  provide a 
wealth of information about its dynamics.

A very fruitful approach towards unraveling the mysteries of M-theory
is provided by the matrix formulation initiated in \cite{banks}, known as 
`Matrix theory.' Its basic postulate is that the full eleven-dimensional 
dynamics of M-theory can be captured by means of the many-body 
quantum mechanics of $N$ D-particles of IIA string theory, in the limit 
where $N$ is taken to infinity. From the eleven dimensional perspective,
this limit can be interpreted as applying a Lorentz boost, so that all degrees 
of freedom end up with a very large momentum along the extra eleventh direction. 

One can formally combine this limiting procedure with a 
decompactification limit of the eleventh direction, so that the IIA string 
and M-theory compactification manifolds become identified.  In this way, 
a fundamentally new perspective on non-perturbative string theory arises,
in which its complete particle spectrum in a given dimension becomes identified 
with a relatively convenient subset of states compactified to one dimension less. 
Though in essence a light-cone gauge description, Matrix theory thus manages to 
provide a rather detailed theoretical framework for exploring the microscopic 
dynamics of  M-theory, while preserving many of the geometrical foundations of 
string theory.  There is now indeed a rather well-developed understanding of 
how Matrix theory incorporates all perturbative degrees of freedom and 
interactions of type II strings \cite{motl} \cite{tomnati} 
\cite{matrix-string}. 

In these notes we will review some of the recent developments in this
approach to M-theory. In particular we will discuss
in detail the recently proposed formulation of M-theory compactified
on $T^5$, based on the 5+1-dimensional world-volume theory of $N$ NS
5-branes \cite{nati}. We will pay special attention to the appearance
of the M5-brane in this proposal, and with the aim of illuminating
the direct correspondence between the framework of \cite{nati} 
and our earlier studies \cite{fivebrane}. In addition we will highlight 
the main ingredients of Matrix string theory \cite{matrix-string},
and discuss its application to the study of the interacting string-like
theory on the 5-brane worldvolume.

\medskip

\section{U-duality and the M5-brane}

\noindent
The degrees of freedom and interactions of M-theory compactified 
on a five-torus $T^5$ provide a representation of the
maximal central extension of the 5+1 dimensional ${\cal N} = 4$
supersymmetry algebra, which in a light-cone frame takes the 
form\footnote{Hence $\alpha$ and $\beta$ are spinor indices of the 
transversal $SO(4)$ rotation group.}
\ba
\label{susy}
\left\{\widetilde{Q}^a_\alpha , \widetilde{Q}^b_\beta \right\} \is p_+
\delta_{\alpha\beta} \ww^{ab} \nonumber\\ \left\{Q^a_\alpha ,
\widetilde{Q}^b_\beta\right\} \is p_I \gamma^I_{\a\b} \ww^{ab} +
\delta_{\a\b}Z^{ab}\\ \left\{{Q}^a_\alpha , Q^b_\beta \right\} \is
p_- \delta_{\alpha\beta} \ww^{ab}\nonumber
\ea
Here the internal labels $a,b = 1, \ldots, 4$ are $SO(5)$ spinor
indices, and $I=1,\ldots, 4$ label the four transverse 
uncompactified space dimensions.  
The central charge matrix $Z^{ab}$ comprises a total of 16 internal
charges, which split up in five quantized Kaluza-Klein 
momenta $p_i$ with $i=1, \ldots, 5$, ten charges $m_{ij}= -m_{ji}$ 
that specify the winding numbers of the M membranes around the 
five-torus, and one single charge $\qf$ that represents the wrapping 
number of the M5 brane. To make this interpretation more 
manifest, it is often convenient to expand $Z_{ab}$ in terms of 
$SO(5)$ gamma matrices as 
\be
\label{decomp}
Z_{ab}= \qf \ww_{ab} + p_i \gamma^i_{ab} + m_{ij} \gamma^{ij}_{ab}.
\ee 
More generally, the decomposition of $Z_{ab}$ into integral
charges depends on the moduli of the $T^5$ compactification, which
parametrize the coset manifold 
\be
\label{calm}
{\cal M} =  SO(5,5)/SO(5)\times SO(5).
\ee
The U-duality group is the subgroup of $SO(5,5)$ rotations that 
map the lattice of integral charges on to itself, and is thus 
identified with 
\be
\label{calu}
{\cal U} =SO(5,5,\Z).
\ee
{}From the eleven dimensional viewpoint, the moduli space (\ref{calm})
is parametrized by the metric $G_{mn}$ and three-form potential $C_{mnk}$ 
defined on the internal manifold $T^5$. After replacing the 3-form $C$ 
on $T^5$ by its Hodge-dual 2-form $B =*C$, this parametrization of ${\cal M}$
reduces to the familiar one from the toroidal compactification of 
type II string theory on $T^5$. This correspondence, which is strengthened
by the fact that the U-duality group (\ref{calu}) takes the form
of the $T^5$ T-duality group, has become a central theme in recent 
attempts to formulate the $T^5$ compactification of M-theory 
\cite{fivebrane,nati}.

To study the spectrum of M-theory on $T^5$, it is natural to start
from the maximal dimensional object that can wrap around the
internal manifold, which is the M5 brane. {}From its description as 
a soliton it is known that its low energy collective modes form a tensor 
multiplet of the ${\cal N}=(2,0)$ world-brane supersymmetry, consisting 
of five scalars, an anti-symmetric tensor with self-dual field strength 
$T=dU$ and 4 chiral fermions. As shown in \cite{fivebrane}, this worldvolume 
field theory can be extended to include all the above 16 charges of 
M-theory on $T^5$ via an appropriate identification of flux configurations.  

To make this concrete, consider a M5-brane with topology of 
$T^5\times \R$ where $\R$ represents the world-brane time $\tau$.  
To begin with, we can include configurations carrying a non-zero
flux 
\be
\label{mflux}
m_{ij} = \int_{T^3_{ij}} dU 
\ee
through the 10 independent three-cycles $T^3_{ij}$ in $T^5$. An M5-brane
for which these charges are non-zero represents a bound state with
a corresponding number of membranes wrapped around the dual $2$-cycle. 
The charges $m_{ij}$ are part of a 16-dimensional spinor
representation of the U-duality group $SO(5,5,\Z)$, which
is isomorphic to the odd (co)homology of the five-torus $T^5$. 
This observation motivates us to try to write the other charges
$\qf$ and $p_i$ as fluxes of 5- and 1-form field strengths through the
5-cycle and 1-cycles on $T^5$. 

Before presenting this construction, let us mention that
in course we will in fact arrive at a rather new notion of the
M5-brane worldvolume theory, that is rather different
from the usual one in terms of the physical location of the fivebrane
soliton in space-time.  In particular, the M5-brane worldvolume theory
we will arrive at lives on an auxiliary 5+1-dimensional parameter
space, denoted by $\widetilde{T}^5$. This auxiliary space is only
indirectly related to the embedded M5-brane soliton in physical
space-time, via an appropriate geometrical interpretation of the
worldvolume fields.

In this new set-up, we can introduce a winding number of the M5-brane via
\be
\label{qvolume}
\qf=\int_{\widetilde{T}^5} dV,
\ee
where the 5-form $dV$ is interpreted as the pull-back to the 
auxiliary torus $\widetilde{T}^5$ of the constant volume element on the 
target $T^5$. We would like to turn the 4-form potential $V$ into an 
independent field, that is part of the world-brane theory. 

To this end we imagine that the (2,0) world-brane tensor multiplet 
in fact provides an effective {\it light-cone description} of the 
M5-brane. Concretely this means that of the five scalars, we will
interpret only four as describing transversal coordinates in 
space-time. This leaves us with one additional scalar field $Y$. 
Since the world-brane is 5+1-dimensional, this scalar can be 
dualized to a 4-form, which we will identify with the 4-form $V$ 
used in (\ref{qvolume}). Hence we will assume the 5-form field 
strength $W=dV$ associated with $V$ is normalized such that the 
associated flux is a non-negative integer.  

Quite generally, in a light-cone formalism for extended objects one 
typically obtains a residual gauge symmetry under volume preserving 
diffeomorphisms \cite{hoppe}. For the case of the 5-brane these can be used to 
eliminate all dependence on the five compact embedding coordinates 
$X^i$  (that map the auxiliary $\widetilde{T}^5$ into the target $T^5$), 
except for the volume-form 
\be
dV\!=\!dX^i\!\wedge\! dX^j \!\wedge\!dX^k \!\wedge\!
dX^l\!\wedge\!  dX^m\epsilon_{ijklm}.
\ee 
The five spatial components of $*V^i$ can roughly be identified 
with the embedding coordinates $X^i$.  In this correspondence
the linearized gauge-transformation $V\ra V+d\Lambda$ of the 4-form 
potential describes a volume preserving diffeomorphism, since it leaves 
$dV$ invariant.
 
We can formalize the duality transformation between $V$ and $Y$ by
taking $W$ to be an independent 5-form and introducing $Y$ as the
Lagrange multiplier that imposes the Bianchi identity $dW=0$. The fact
that $W$ has integral fluxes implies that $Y$ must be a periodic
field, \ie $Y \equiv Y+2\pi r$ with r integer. The 5 remaining charges
$p_i$ can thus be identified with the integer winding numbers of $Y$
around the 5 one-cycles
\be
\label{momenta}
p_i = 
\oint_{T^1_i} d Y. 
\ee
Since on-shell $\Pi_V \equiv \delta S/\delta\dot{V} = dY$, we deduce 
that $dY$ is the canonical momentum for $V$. From this it is 
straightforward to show that these operators are indeed the generators 
of translations along the internal directions on the target space
5-torus. The formula for the M5-brane wrapping number now becomes 
\be
\label{qfive}
\qf= \int_{\widetilde{T}^5} \Pi_Y,
\ee 
with $\Pi_Y$ the conjugate momentum to $Y$.  

In this way all the 16
charges $(q,p_i,m_{ij})$ have been written as fluxes through the odd
homology cycles on the M5-brane.  These 16 bosonic zero modes
form the world-volume superpartners of the 16 fermionic zero modes,
that represent the space-time supersymmetries that are broken by the
M5-brane soliton. Based on this result, it was shown in \cite{fivebrane} 
that the (2,0) worldvolume theory of
the M5-brane could be used to represent the full
maximally extended space-time supersymmetry algebra (\ref{susy}).

The distinction between the above and the standard
notion of the 5-brane world-volume is perhaps best explained
in analogy with the worldsheet description of toroidally
compactified string theory. According to the strict definition,
the worldsheet of a string refers to its particular
location in space-time, and thus also specifies a given 
winding sector.  For strings, however, it is well known to be
advantageous to combine into the worldsheet CFT all superselection 
sectors corresponding to all possible momenta and winding numbers. 
T-duality for example appears as a true symmetry of the worldsheet 
CFT only if it includes all sectors. The notion of the M5 worldbrane 
theory used here is indeed the direct analogue of this: just as
for  strings, the various wrapping numbers and momenta appear 
as quantized zero modes of the worldvolume fields. A second 
important analogy is that, just like the CFT Hilbert space 
includes unwrapped string states, we can in the above set-up 
also allow for unwrapped M5 brane states with $\qf=0$.

This naturally leads to the further suggestion that it should be
possible to extend the M5 worldbrane theory in such a way that 
U-duality becomes a true symmetry (just like T-duality in 
toroidal CFT). In \cite{fivebrane} it was shown that this
can indeed be achieved by assuming that the worldvolume theory on
$\widetilde{T}^5$ is described by a 5+1-dimensional second 
quantized string theory, whose low energy effective modes 
describe the (2,0) tensor multiplet. Via the $SO(5,5,\Z)$ T-duality 
of this ``micro-string theory'', this produces a manifestly 
U-duality invariant description of the BPS states of all
M-theory compactification down to 6 or more dimensions
\cite{fivebrane}.

Until quite recently, only limited theoretical tools were available to
analyze the detailed properties of this conjectured micro-string
theory, or even to confirm its existence. The traditional methods
usually start from the physical world volume of a macroscopic M5-brane
soliton, and the micro-string theory is in fact not easily visible
from this perspective. As seen e.g. from the identifications
(\ref{momenta}) and (\ref{qfive}), the local physics on the target
space torus $T^5$ is somewhat indirectly related to that on the
parameter space $\widetilde{T}^5$ where the micro-string lives.  The
specification of a particular winding sector $\qf$, for instance,
breaks the U-duality group to geometric symmetry group $SL(5,\Z)$, and
thus hides the stringy part of the micro-string T-duality
symmetry. This illustrates that it may be misleading to think about
the micro-strings directly as physical strings moving on the
space-time location of the M5-brane soliton.

\section{BPS Spectrum}

Further useful information about the relation between the 
auxiliary and target space can be obtained by considering the masses 
of the various BPS states. 

The world-brane theory carries a chiral ${\cal N}\! = \! 2$ 
supersymmetry algebra. To write this algebra, it is 
convenient to introduce chiral $SO(4)$ spinor indices 
$\a$ and $\dot{\alpha}$ for the transverse rotation group 
(which represents an R-symmetry of the world-volume supersymmetry). 
We then have
\ba
\label{wsusy}
\left\{{Q}^\a_a , Q^\b_b \right\} \is
\epsilon^{\a\b} (\gamma^0_{ab} H +\gamma^i_{ab} (P_i + W_i)), \nonu
\left\{{Q}^{\dot\alpha}_a , {Q}^{\dot\beta}_b \right\} 
\is \epsilon^{\dot\a \dot\b} (\gamma^0_{ab} H +
\gamma^i_{ab} (P_i-W_i)) .
\ea
Here $H$ is the world-volume Hamiltonian,
and $P_i$ is the integrated momentum flux through the five-torus
\be
H = \int_{\widetilde{T}^5}\!\!  
T_{00},  \qquad P_i = \int_{\widetilde{T}^5}\!\! T_{0i},
\ee
where $T_{mn}$ denotes the world-volume energy-momentum tensor. 
Hence $H$ and $P_i$ represent the generators of time and space
translation on the 5+1-dimensional parameter space-time.
The operators $W_i$ that appear in the algebra 
(\ref{wsusy}) represent a vector-like central charge, and is a 
reflection of the (possible) presence of string-like objects
in the world-volume theory. This vector charge $W_i$
indeed naturally combines with the momenta $P_i$ into a 10 
component vector representation of the $SO(5,5,\Z)$ world-volume 
T-duality group. 

The world-brane supercharges represent the unbroken part of 
space-time supersymmetry algebra. The other generators must
be identified with fermionic zero-modes, that are the world-brane
superpartners of the bosonic fluxes described above. In order
to exactly reproduce the 5+1-dimensional space-time supersymmetry 
(\ref{susy}), however, we must project onto states for which
the total momentum fluxes $P_i$ and vector central charges $W_i$
all vanish
\be
\label{levelmatch}
P_i \, = \, 0, \qquad \quad W_i \, =\, 0.
\ee
These are the analogues of the usual level matching relations
of light-cone string theory. In addition, we must impose
\be
H =p_-
\ee
which reflects the identification of the world-volume time coordinate
$\tau$ with the space-time light-cone coordinate $x^+$.
The mass-shell condition $p_+p_- = m^2$ (we assume for simplicity
that all transverse space-time momenta $p_I$ are zero) 
relates the space-time mass of M5-brane 
configuration to the world-volume energy $H$ via
\be
\label{mass-shell}
m^2 = p_+ \, H
\ee
This relation, together with identification of the central charge
$Z_{ab}$ as world-brane fluxes, uniquely specifies the geometric 
correspondence between the auxiliary torus $\widetilde{T}^5$ and the 
target space torus $T^5$. To make this translation explicit, let 
us denote the lengths of the five sides of each torus by 
$\widetilde{L}_i$ and $L_i$ respectively, and let us fix
the normalization of the fields $dU$ and $dY$ in accordance
with the flux quantization formulas (\ref{mflux}), (\ref{momenta}),
and (\ref{qfive}). Since the left-hand sides are all integers,
this in particular fixes the dimensionalities of the two-form 
$U$ and scalar $Y$.

To simplify the following formulas, we will further use {\it natural units}
both in the M-theory target space, as well as on the auxiliary space.
In the target space, these natural units are set by the
eleven dimensional Planck scale $\ell_p$, while on the world-volume,
the natural units are provided by the micro-string length scale 
$\widetilde{\ell}_s$. Apart from the size of the five torus $\tilde{T}^5$,
this string length $\widetilde{\ell}_s$ is indeed the only other length
scale present in the 5+1-dimensional world-volume theory.

\newcommand{\YM}{{{}_{Y\! M}}}

Now let us consider the physical masses of the various 
1/2 BPS states of M-theory. These states
can carry either pure KK momentum charge, or pure 
(non-intersecting) membrane and fivebrane wrapping number. 
From the target-space perspective, we deduce that 
these states have the following masses (in 11-dimensional
Planck units $\ell_p = 1$)
\ba
\label{Mass}
\hbox{\it KK momenta:} & & \quad
m^2 \, = \, ({p_i/ L_i} )^2 
\nonumber \\[2.7mm]
\hbox{\it membranes:} & & \quad
m^2 \, = \, 
(m_{ij} L_i L_j)^2 
\\[2.7mm]
\hbox{\it fivebranes:} & & \quad
 m^2 \, = \, 
( \, \qf \, V_5\, )^2 
\nonumber
\ea
where $V_5 =L_1\ldots L_5$ denotes the volume of the M-theory five 
torus $T^5$ in Planckian units.

In the world-volume language these 1/2 BPS states all correspond 
to the supersymmetric ground states  that carry a single flux quantum 
number. Their mass is therefore determined via the mass-shell
relation (\ref{mass-shell}) by 
the world-volume energy carried by the corresponding flux
configuration. To deduce this energy, it is sufficient to note that
at very long distances, the world-volume theory reduces to a 
non-interacting free field theory of the (2,0) tensor multiplet. 
Hence the flux contribution to the M5-brane Hamiltonian can be
extracted from the free field expression\footnote{Here we have
used that the self-duality relation $\Pi_U = dU$. We further left 
out the transverse scalars $X_I$ since their fluxes are set to zero.}
\be
\label{haha}
H =  {1\over \beta} \int_{\widetilde T^5} \! \! d^5 \! x 
\left[ 
|dY|^2 + |dU|^2 + 
|\Pi_Y|^2 \right]
\ee
where $\beta$ is an arbitrary parameter (to be related later 
on to the light-cone momentum $p_+$ in space-time). 
Now using (\ref{mflux}), (\ref{momenta}), and (\ref{qfive}), we deduce
that the 1/2 BPS states have the following worldvolume energies (in
micro-string units $\widetilde{\ell}_s = 1$)
\ba
\hbox{\it KK momenta:} & & \quad
\, {\widetilde V_5\over \beta } 
\Bigl({p_i\over\widetilde L_i}\Bigr)^2 
\nonu
\hbox{\it membranes:} & & \quad
H \, = \, {(m_{ij} \widetilde L_i \widetilde L_j)^2 \over 
\beta \widetilde V_5}
\\[2mm]
\hbox{\it fivebranes:} & & \quad
H \, = \, 
{
\qf^2\over \beta \widetilde V_5} 
\nonumber
\ea
Here $\widetilde{V}_5 =\widetilde{L}_1\ldots\widetilde{L}_5$ 
denotes the volume of the auxiliary space $\widetilde{T}^5$
in string units.

From the condition that all these three expressions reproduce 
the correct mass formulas (\ref{Mass}) via (\ref{mass-shell}), 
we can now read off the M-theory parameters in terms of the
world-volume parameters. First of all we see that the two 
five tori $T^5$ and $\widetilde{T}^5$ in fact have the same
{\it shape}, \ie they are related by means of an overall
rescaling
\be
L_i = v \widetilde{L}_i 
\ee
with $v$ some dimensionless number. The remaining M-theory
parameters we need to determine are $V_5$ and $p_+$, and we need to express
them in terms of $\widetilde{V}_5$ and $\beta$ in (\ref{haha}). 
A little algebra show that 
\be
V_5 = {1\over \widetilde{V}_5^{{2/ 3}}}, \qquad \quad 
p_+ =  {\beta \over \widetilde{V}_5^{1/ 3}}.
\ee
In particular we find that the wordvolume radii are related to
the M-theory radii by
\be
\widetilde{L}_i = {L_i \over V_5^{1/2}}.
\label{radii}
\ee
Hence we see in particular that the decompactification limit of M-theory
corresponds to the zero volume limit of the auxiliary torus. 

Of course, here we have to take into account that T-duality of the
micro-string theory can relate different limits of the
$\widetilde{T}^5$ torus geometry. This fact plays an important role
when we consider partial decompactifications. For example, if we send
one of the radii of the M-theory five-torus, say $L_5$, to infinity,
then (\ref{radii}) tells us that, after a T-duality on the remaining
four compact directions, the worldvolume of $\widetilde{T}^5$ tends to
infinity.  In this way we deduce that the BPS states of the $T^4$ 
compactification of M-theory is described by the zero slope limit of 
the micro-string theory on a $T^5$ world-volume. This is of importance 
in understanding the $SL(5,\Z)$ U-duality group of M-theory 
compactifications on $T^4$ \cite{rozali,brs}. Indeed, all U-duality
symmetries for $T^d$ compactifications with $d\leq 5$ can be understood
in this way as geometric symmetries of the micro-string model.

Finally, let us mention that as soon as one considers states with more
general fluxes, they are at most 1/4 BPS and the mass formula becomes
more involved.  One finds
\be
\noindent
m\bps^2 = m_0^2 +
2 V_5 \sqrt{({\kK_i/ L_i})^2 + {\textstyle{1\over V^2_5}}(\wW_i L_i)^2},
\ee
with
\ba
\label{kK}
m_0^2 \is (\qf V_5)^2 + (p_i/L_i)^2 + (m_{ij}L_iL_j)^2, \nonu
\kK_i \is  \qf\,p_i+ \hf \epsilon_{ijklm}m_{jk}m_{lm}, \\[2.5mm]
\wW_i \is  m_{ij} p_j. \nonumber
\ea
These formulas are completely duality invariant,  and  were
reproduced via a world-volume analysis in \cite{fivebrane}. 
 A crucial new ingredient in the description of BPS states
with a non-zero value for $\kK_i$ is that they necessarily contain worldbrane
excitations with non-zero momentum on $\widetilde{T}^5$,\footnote{This is
implied by the level-matching condition $P_i=0$, since $\kK_i$ in fact
represents the zero-mode contribution to $P_i$. Notice further that for
$\qf\neq 0$, $\kK_i$ contains a contribution proportional to $\qf$ 
times the space-time KK momentum $p_i$, which indeed reflects the 
geometric relation between world-volume and target space translations.} 
while $\wW_i\neq 0$ necessarily describes states with non-zero
winding number for the micro-strings. Hence, without the micro-string
degrees of freedom, the M5-brane would have had no BPS-states with
non-zero value for $\wW_i$. The geometrical interpretation of the
bilinear expressions for $\kK_i$ and $\wW_i$ will become more clear
in the following sections.

\section{M-theory from NS 5-branes} 

The above attempt to unify the complete particle spectrum of 
a given M-theory compactification in terms of one single world-volume 
theory has now found a natural place and a far reaching generalization 
in the form of the matrix formulation of M-theory initiated 
in \cite{banks}. While the previous sections were in part based  
on an inspired guess, motivated by U-duality and the form of 
the space-time supersymmetry algebra, in the following sections
we will describe how the
exact same structure can be derived from the recently proposed
Matrix description \cite{nati} of M-theory on $T^5$.
So we will again start from the beginning.

According to the Matrix theory conjecture, one can
represent all degrees of freedom of M-theory via an appropriate
large $N$ limit of the matrix quantum mechanics of the D0-branes
\cite{banks}. The true extent this proposal becomes most visible
when one combines this large $N$ limit with a decompactification
limit $R\! \ra \!\infty$ of the 11-th M-theory direction. The ratio
\be
p_+ = {N/ R}
\ee
represents the light-cone momentum in the new decompactified
direction.

U-duality invariance implies that any 
charge of compactified IIA string theory can in principle be 
taken to correspond to the eleventh KK momentum $N$.  
One can thus reformulate the original matrix
conjecture of \cite{banks} by applying some duality symmetry of the
type II string to map the D0-branes to some other type of branes.
Compactification of matrix theory indeed generically proceeds by using
the maximal T-duality to interchange the D0 brane charge by that of
the maximal dimensional D p-brane that can wrap around the internal
manifold. In our specific example of $T^5$ this would be the
D5-brane. Alternatively, as was first suggested in \cite{nati} (following
up earlier work \cite{rozali} \cite{brs} \cite{fivebrane}), one
can choose to identify this charge $N$ with the NS five-brane wrapping
number around the five-torus $T^5$.  

In the following table we have indicated the chain of duality mappings 
that provides the dictionary between the original matrix set-up
of \cite{banks} and the dual language of \cite{nati}:
\begin{eqnarray}
\renewcommand{\arraystretch}{1.5}
\begin{array}{cccccccc}  
\begin{array}{|l|}
  \hline
  N \\[2mm]
  \hline
  p_i\\[2mm]
  \hline
  m_{ij}\\[2mm]
  \hline
  \qf\\[2mm]
  \hline
  \end{array}
  &= &
  \begin{array}{|l|}
  \hline
  D0\\[2mm]
  \hline
  \KK\\[2mm]
  \hline
  D2\\[2mm]
  \hline
  \NS 5\\[2mm]
  \hline
  \end{array}
  & \mapright{T_5} &
  \begin{array}{|l|}
  \hline
  D5\\[2mm]
  \hline
  \NS 1\\[2mm]
  \hline
  D3\\[2mm]
  \hline
  \NS 5\\[2mm]
  \hline
  \end{array}
  &\mapright{S} &
 \\[1.5cm]
\mbox{M on $T^5$} && \mbox{IIA on $T^5$} && \mbox{IIB on $\widehat{T}^5$}&&
\end{array}\nonumber
\renewcommand{\arraystretch}{1.0}
\end{eqnarray}

\begin{eqnarray}
\renewcommand{\arraystretch}{1.5}
\begin{array}{cccccc}  
 \qquad   & \mapright{S} &       
  \begin{array}{|l|}
  \hline
  \NS 5\\[2mm]
  \hline
  D1\\[2mm]
  \hline
  D3\\[2mm]
  \hline
  D5\\[2mm]
  \hline
  \end{array}
  &\mapright{T_5} &
  \begin{array}{|l|}
  \hline
  \NS 5\\[2mm]
  \hline
  D4\\[2mm]
  \hline
  D2\\[2mm]
  \hline
  D0\\[2mm]
  \hline
 \end{array} \\[1.5cm]
&& \mbox{IIB on $\widehat{T}^5$} && \mbox{IIA on $\tilde{T}^5$}
\end{array}\nonumber
\renewcommand{\arraystretch}{1.0}
\end{eqnarray}
\noindent

\parbox{6.6cm}{\it 
Table 1. \ Interpretation of the quantum
numbers and their translation under the
duality chain.} 

\medskip

\noindent
We notice that all 16 quantized charges, that make up the
central charge matrix $Z_{ab}$, have become D-brane wrapping numbers
in the last two columns, and furthermore that the $SO(5,5,\Z)$
U-duality symmetry is mapped onto the (manifest) T-duality symmetry of
the type II theory. Note further that the D-brane charge vector indeed 
transforms in a spinor representation of the T-duality group.

The string coupling, string scale and dimensions of the
five torus are related through this duality chain as 
follows.\footnote{For this
section we benefited from valuable discussions with G. Moore. A
similar presentation can indeed be found in \cite{candy}.} The chain 
consists of a $T_5$ duality on all five direction of the $T^5$, an 
S-duality of the IIB theory, and finally again a maximal T-duality 
on $T^5$.  The end result is a mapping from large $N$ D-particle 
quantum dynamics to that of $N$ NS five-branes in IIA theory.

On the left, we have started with the conventional definition of
M-theory compactified on a five-torus $T^5$, with dimensions $L_i$,
$i=1,\ldots,5$, as the strong coupling limit of type IIA
string theory on the same torus $T^5$. 
Under the first $T_5$ duality, a D0-brane becomes the D5-brane
wrapped around the dual torus $\widehat{T}_5$ with dimensions
\be
\Sigma_i= {\ell_s^2 \over  L_i} = {1\over R L_i}.
\ee
Here we used (\ref{scales}), and again work in eleven dimensional
Planck units $\ell_p = 1$. 
Following the usual rule of $T$-duality, the string coupling $g_s$  
gets a factor of the volume in string units, so the
dual coupling $\widehat{g}_s$ becomes (see again (\ref{scales})
\be
\widehat{g}_s= {g_s \ell_s^5 \over V_5}= {1\over R V_5}
\ee
where $V_5$ is again the five-volume of $T^5$ (in Planck units).
The subsequent S-duality transformation maps the D5-brane onto the IIB  
NS5-brane, inverts the string coupling, and changes the string scale  
by a factor of $\sqrt{\widehat{g}_s}$. So we have
\be
\label{ltilde}
\tilde{g}^\prime_s=V_5 R,\qquad \tilde{\ell}^2_s=  
{1\over R^2 V_5}.
\ee
The size of the five torus of course remains unchanged under the 
S-duality. 

Finally we perform the $T_5$ duality again, so that $N$ counts 
IIA fivebrane charge. The dimensions of the resulting five-torus
$\tilde{T}^5$ are 
\be
\label{reln}
\tilde{L}_i = {L_i\over R V_5}.
\ee
while the string coupling $\widetilde{g}_s$ of the IIA  
string after the $T_5ST_5$ transformation becomes 
\be
\label{dualcoupling}
\tilde{g}^2_s = 
{R^2\over V_5}
\ee
The string scale of the IIA theory is the same as in (\ref{ltilde}).

In \cite{nati} convincing arguments were given for the existence
of a well-defined limit of type II string theory, that isolates the
world volume dynamics of $N$ NS five-branes from the 9+1-dimensional
bulk degrees of freedom. The key observation made in \cite{nati}
is that, even in the decoupling limit $\tilde{g}_s \ra 0$ for
the bulk string interactions, one is still left with a non-trivial
interacting 5+1 dimensional theory that survives on the world volume
of the 5-branes. This theory depends on only two dimensionless 
parameters, namely $N$ and the size of the torus $\tilde{T}^5$
in units of the string length $\tilde{\ell}_s$. If one keeps the last
ratio finite, the system will contain string like degrees of freedom.
According to the proposal put forward in \cite{nati}, 
this world volume theory of $N$ NS 5-branes in the limit
of large $N$ may provide a complete matrix theoretic description 
of M-theory compactified on $T^5$.

This proposal in fact needs some explanation, since as seen from 
(\ref{dualcoupling}) this decoupling limit $\widetilde{g}_s \ra 0$ in
fact contradicts the decompactification limit $R\ra \infty$
that is needed for obtaining M-theory on $T^5$. The resolution
of this apparent contradiction should come from the particular
properties of the world-volume theory of $N$ 5-branes for
large $N$. As we will see later, via a mechanism familiar e.g. 
from studies of black holes in string theory \cite{maldacena}
or matrix string theory \cite{motl}\cite{tomnati}\cite{matrix-string}, 
the system of $N$ wrapped NS 5-branes contains a low energy sector 
with energies of order $1/N$ times smaller than the inverse size of the 
five-torus $\tilde{T}^5$. These low energy modes are the ones that 
will become the M-theory degrees of freedom, and the large $N$ limit 
must be taken in such a way that only these states survive.
This implies that the typical length scale of the relevant
configurations is larger by a factor of $N$ than 
the size of $\tilde{T}^5$. 

When one analyzes this low lying energy spectrum   
for very large $N$, one discovers that it becomes essentially 
independent of the string coupling 
constant $\widetilde{g}_s$, provided $\widetilde{g_s}$ does not
grow too fast with $N$. It thus seems reasonable to assume
that in an appropriate large $N$ limit, the world-volume
theory of the NS 5-branes itself becomes independent of 
$\widetilde{g}_s$.  This opens up the possibility that
the weak coupling limit that was used in \cite{nati} for
proving the existence of the 5+1-dimensional worldbrane theory, 
in fact coincides with the strong coupling regime that  
corresponds with the decompactification limit of M-theory. 

Before we continue, let us remark that this Matrix theoretic set-up 
can already be seen to reproduce many qualitative and quantitative
features of the framework described in the previous sections.
Evidently, the world-volume theory of the $N$ NS 5-branes plays
a very analogous role to the proposed M5-brane world-volume theory
of sections 2 and 3. Most importantly, in both set-ups the U-duality
symmetry of M-theory becomes identified with a T-duality of a
5+1-dimensional worldvolume string theory. In addition, as we will 
see a little later, all 16 charges of M-theory arise in the present Matrix 
theory set-up \cite{nati} via the 
exact same identification of fluxes as described in section 2.

Finally, by comparing the formulas (\ref{ltilde}) and (\ref{reln}) 
with those of the previous section, we see that the size and shapes 
of the respective auxiliary five-tori $\widetilde{T}^5$ and $\tilde{T}^5$ 
(notice however the subtle difference in notation) are expressed via an
identical set of relations in terms of those the M-theory torus $T^5$. 
The micro-string torus $\widetilde{T}^5$ of sections 2 and 3 is 
indeed related the Matrix theory torus $\tilde{T}^5$ used above via
a rescaling with a factor $N$. Since (as will be 
elaborated later on) the micro-string scale $\widetilde{\ell}_s$ 
is also $N$ times larger than the fundamental string scale, the two
auxiliary five tori have the same size relative to the respective
string scales. This fact ensures that the standard T-duality 
symmetry of the small $\tilde{T}^5$ indeed translates into an 
isomorphic T-duality group that acts on the large $\widetilde{T}^5$ 
on which the micro-string lives.

Via this formulation \cite{nati}, Matrix theory thus naturally incorporates
our earlier ideas on the M5-brane. More importantly, however, it also 
provides a concrete prescription for including interactions, both among the
M5-branes as well as for the micro-string theory inside its world-volume.

\section{SYM theory on $T^5$}

One possible starting point for concretizing this matrix description
of M-theory on $T^5$ is provided by the large $N$ limit of
5+1-dimensional super Yang-Mills theory
\ba
\label{susym}
S \is \frac{1}{g^2}\int \! d^6 \! x\, {\rm tr}\, \Bigl(\frac{1}{4}
F_{\mu\nu}^2+\frac{1}{2}(D_{\mu}X^I)^2  \nonu
& & \ +
\psi\Gamma^i D_i \psi
+\psi\Gamma^I[X^I,\psi]+\frac{1}{4}[X^I,X^J]^2 \Bigr)
\nonumber
\ea
Here $I = 1, \ldots, 4$ labels the 4 Higgs scalar fields. This theory can be 
thought of as describing the low energy collective modes of $N$ D5 branes 
of IIB string theory (the second column in the above table), wrapped around
the dual five-torus $\hat{T}^5$. In terms of the parameters above, the
Yang-Mills coupling of the D5-brane world-volume theory is then
\be
\label{ymcoupling}
{1\over g^2_{\YM}} = {1\over \widehat{g}_s\ell_s^2}= V_5R^2
\ee 
Alternatively, we can apply an S-duality and view the large $N$ SYM
model as the low energy world volume theory of $N$ IIB NS 5-branes.
Note that the above YM coupling then becomes equal to the
fundamental NS string tension $g_{\YM}=\widetilde{\ell}_s$.  

According
to the prescription of Matrix theory, finite energy states of M-theory
on $T^5$ are obtained by restricting to states of the SYM model with
energies of order $1/N$. Since these states are all supported at
scales much larger than the string scale, it seems an allowed
approximation to discard the stringy corrections to the $U(N)$ SYM
description of the NS 5-brane dynamics. 
An obvious subtlety with making this low-energy truncation
is that the 5+1 dimensional SYM model
in itself does not represent a well-defined 
renormalizable field theory. It is conceivable, however, that via 
the combined large $N$ and IR limit that is needed for Matrix theory, 
one can still extract well-defined dynamical rules for the subset 
of the SYM degrees of freedom that are relevant for the description 
of M-theory on $T^5$. 

A second remark is that the Yang-Mills description in fact
seems incomplete. Namely, to describe all possible quantum
numbers, one should be able to include all possible
bound states of the NS 5-branes with the various D-branes.
Generally these are described by turning on appropriate
fluxes of the SYM theory. For the D1-branes and D3-branes,
the corresponding winding number are represented via the electric and
magnetic flux quantum numbers
\be
\label{flux}
p_i =
\int_{T^4_i } \! \tr \, E,\qquad \ 
 m_{ij}=
\int_{T^2_{ij}} \! \tr \, F,
\ee
through the dual 4 and 2-cycles of the $\widehat{T}^5$, respectively.
The D5 brane wrapping number, however, does not seem to appear naturally 
as a flux of the SYM theory. This fact has so far been the main obstacle 
for incorporating the transverse M5 brane into Matrix theory.

Central in the correspondence between the large $N$ SYM theory 
and M-theory on $T^5$ is the fact that (apart from this problem
of the missing $\qf$-charge) the maximally extended supersymmetry 
algebra (\ref{susy}) can be represented in terms of the SYM 
degrees of freedom \cite{branes}. The construction 
is quite standard from the general context of string solitons, 
and relies on the identification of the world-volume supersymmetry 
charges of the SYM model
\ba
\label{unbro}
Q \is
\int \tr \Bigl[ \; \overline{\theta}\, \Bigl(\,
\gamma^{i}E_i + \gamma^I\Pi_I \!
\\[3mm]
& &
-\hf( \gamma^{ij}F_{ij} + \gamma^{iJ} D_i X_J +
\gamma^{IJ}[X_I,X_J] )\Bigr)\Bigr]
\nonumber
\ea
with the generators the unbroken space-time supersymmetries, while the
generators of the broken supertranslations are represented by the fermion
zero modes
\be
\tilde{Q} = \int \! \tr\, \theta.
\ee
The dynamical supersymmetry charges (\ref{unbro}) generate an algebra
of the exact form (\ref{wsusy}), where
\be
\label{pibe}
P_i = \int \tr (E_i F_{ij} + \Pi^I D_i X + \theta^T D_i \theta)
\ee
is the momentum flux and where 
\be
\label{inst}
W_i = \! \int \! \tr  (F\wedge F)_i
\ee
is the instanton winding number.
Hence the SYM field theory explicitly reveals the presence of 
the string like degrees of freedom in the form of the non-abelian 
instanton configurations. 
As we will describe in a later section, 
the long distance dynamics of these instanton strings may 
be used as a quantitative description of the interacting 
micro-string theory.

If $k_i$ denotes the $SU(N)$ Yang-Mills instanton number, 
we evaluate
\be
\label{inst2}
N W_i 
=\hf (m\wedge m)_i + N k_i.
\ee
As we will make more explicit in a moment, the right-hand side can 
indeed be seen to correspond to a total string winding number. 
\footnote{This is true provided the $U(N)$ symmetry is essentially 
unbroken by the Higgs scalars. More generally, if some of the Higgs 
scalars have expectation values that are substantially different from 
zero, $W_i$ will naturally split up into separate 
contributions corresponding to instantons of the smaller unbroken 
subgroups.}
To see this (and to get more insight into how one might recover
the M5-brane from the present set-up) it will be useful to consider 
the world-volume properties of the IIA and IIB NS 5-branes from a 
slightly different perspective.

\section{Winding Microstrings}

{} After the ground breaking work \cite{stromingervafa} on 
the entropy of 5+1-dimensional black holes, it became clear that 
the internal dynamics of the string theory 5-branes
are most effectively described in terms of an effective world-volume 
string theory \cite{fivebrane,maldacena}. These string-like degrees of freedom,
which in the following we will also call micro-strings, 
arise in part due to one-dimensional intersections \cite{stromingervafa}
that occur between D-branes, 
but also due to fundamental NS strings trapped inside NS 5-branes
\cite{sb96}\cite{nati}.
As indicated in \figuurplus{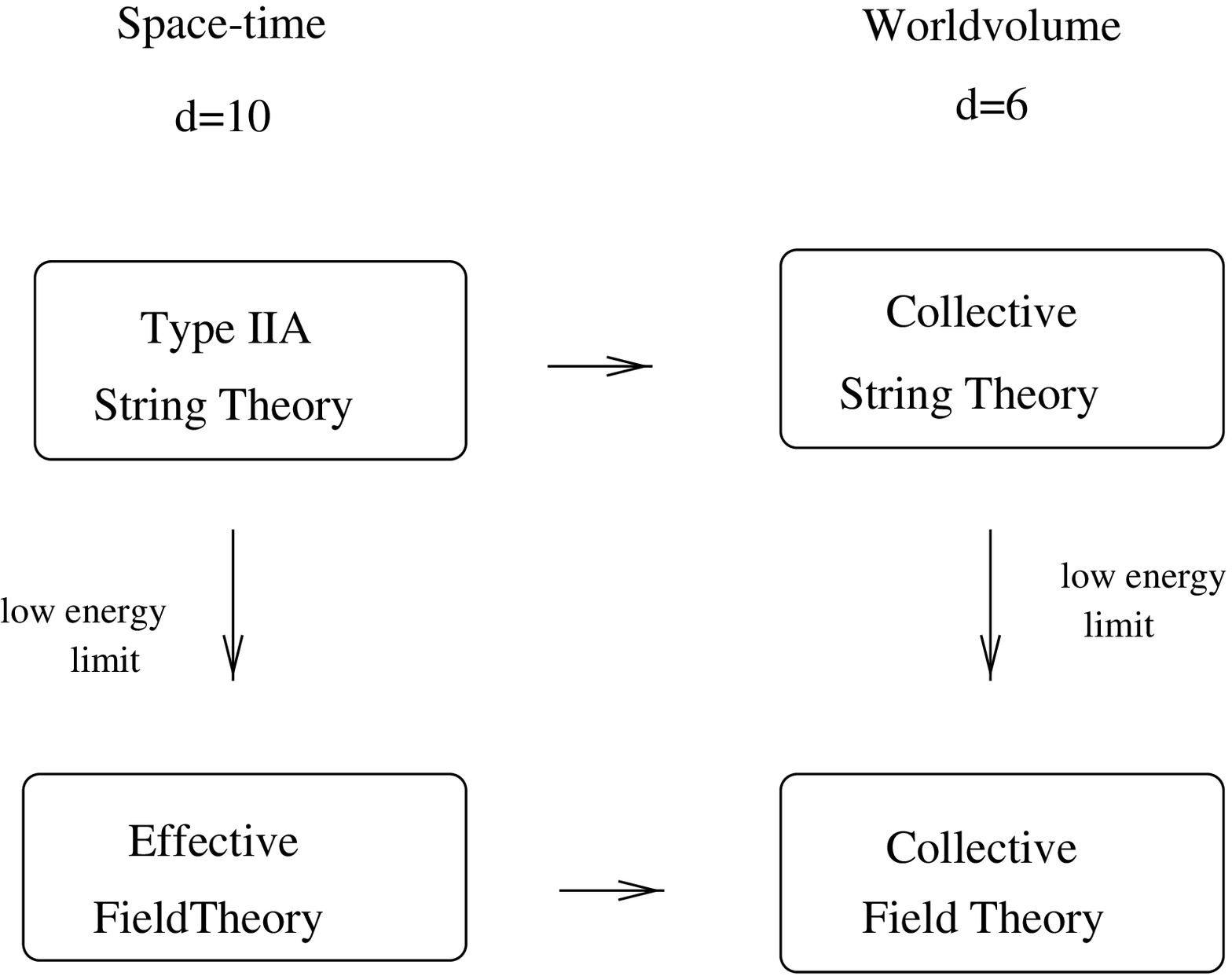}{7cm} {The effective 
world-volume dynamics of the solitonic NS 5-brane 
defined a $d=6$ ``collective string theory'', 
whose low-energy degrees of freedom coincide with the usual 
collective fields. } the relation between the world-volume 
string theory and the low-energy collective modes of the 5-branes 
indeed parallels that between the fundamental type II strings
and the low energy effective field theory in space-time. This
correspondence predicts in particular the precise form of the 
micro-string ground states.

In the past year, this effective string theory has been developed
into a remarkably successful framework for describing the statistical 
and dynamical properties (such as absorption and emission processes) 
of the 5-brane black holes. Until recently, however, it was 
unclear to what extent the corresponding 5+1-dimensional 
string could really be isolated as a separate theory, that exists
independently from the 10-dimensional type II string. The elegant 
argumentation of \cite{nati} provides convincing evidence that this 
is indeed the case. We will now collect some useful topological properties 
of the micro-string theory.

A D-brane in type IIB string theory has odd dimension, and
thus a general configuration of D-branes wound around the internal
manifold $T^5$ will be characterized by a 16 component
vector $\qq^\Bb \in H_{odd}(T^5,\Z)$ in the odd integral
homology of $T^5$. Hence the corresponding D-brane intersection 
strings carry a total winding number given by the 
intersection form on $H_{odd}(T^5,\Z)$, which in terms of our
labeling of charges (see Table 1 in section 3) takes the form 
\be 
\hf (\qq^\Bb\!\! \wedge\! \qq^\Bb)_i \, =\, \qf p_i
+ \hf (m\! \wedge\! m)_i 
\ee 
Fundamental NS strings can also manifest themselves as string-like 
objects within the 5-brane world volume. Inside a bound
state with $N$ NS 5-branes, each NS string can in fact break up into
$N$ separate strings with reduced string tension $1/N\alpha'$.  
These fractional strings are also intersection strings, and string duality 
indeed implies that they are indistinguishable (in the sense of statistics) 
from the above D-brane intersection strings. Denoting the NS string 
winding charge by $k_i$, the total number of intersection 
strings is given by\footnote{Here we assume that all strings and branes 
are at the same location in the non-compact space-time. See also previous
footnote.}
\be 
\label{inter}
N W_i^\B =\hf(m\!  \wedge\! m)_i + \qf
p_i + Nk_i 
\ee
This formula generalizes the formula (\ref{inst2}) for the instanton 
winding charge to the case where $\qf \neq 0$. Notice further the
correspondence with equation (\ref{kK}).

Let us now switch to the IIA perspective. In this case D-branes have 
even dimensions and thus D-brane configurations are characterized by
$\qq^\Aa \in H_{even}(T^5,\Z)$, and string-like intersections
occur between 2- and 4-branes. Thus their total winding number 
is determined via 
\be
\hf (\qq^\Aa\! \! \wedge\! \qq^\Aa)_i \, =\,  m_{ij} p_j
\ee
which is the intersection form on $H_{even}(T^5,\Z)$.
Hence, via the same mechanism as described for the IIB case, the 
total micro-string winding number (including the intersection strings 
between fundamental strings  the $N$ fivebranes) now takes the form
\be
\label{intera}
N W_i^\A = m_{ij} p_j + Nk_i
\ee
where $k_i$ denotes the IIA string winding number. 

As the IIA theory and IIB model are related via T-duality, the two 
micro-string  theories are actually the same, except that the total 
winding numbers $W_i$ get interchanged with the total momentum 
fluxes $P_i$. From the above formulas it is easy to see that the flux 
contribution to e.g. the IIA string winding number $W_i^\A$ is indeed
identical to the flux contribution to the IIB momentum flux $P_i^\B$
given in (\ref{pibe}). A similar check can also be done for the IIB
winding charge $W_i^\B$ and the dual IIA momentum flux $P_i^\A$.

Below we have indicated the Matrix-theory
interpretation the total momentum flux $P_i$ and micro-string 
winding number $W_i$ on the auxiliary $T^5$, as obtained
via the $T_5 \, S \, T_5$ duality chain:

\begin{eqnarray}
\renewcommand{\arraystretch}{1.5}
\begin{array}{cccccccc}
  \begin{array}{|l|}
  \hline
  P_i \\[2mm]
  \hline
  W_i \\[2mm]
  \hline
  \end{array}
 &= &
  \begin{array}{|l|}
  \hline
  \NS 1 \\[2mm]
  \hline
  D4 \\[2mm]
  \hline
  \end{array}
  & \mapright{T^5}  &
  \begin{array}{|l|}
  \hline
  \KK \\[2mm]
  \hline
  D1 \\[2mm]
  \hline
  \end{array}
  &\mapright{S} & 
\\[7mm]
\mbox{M on $T^5$} && \mbox{IIA on $T^5$} && \mbox{IIB on $\widehat{T}^5$}
&& 
\end{array}\nonumber
\renewcommand{\arraystretch}{1.0}
\end{eqnarray}

\begin{eqnarray}
\renewcommand{\arraystretch}{1.5}
\begin{array}{cccccccc}
 \qquad  &\mapright{S} &
  \begin{array}{|l|}
  \hline
  \KK \\[2mm]
  \hline
  \NS 1\\[2mm]
  \hline
  \end{array}
  & \mapright{T^5}  &
  \begin{array}{|l|}
  \hline
  \NS 1\\[2mm]
  \hline
  \KK \\[2mm]
  \hline
 \end{array} \\[7mm]
&& \mbox{IIB on $\widehat{T}^5$} && \mbox{IIA on $\tilde{T}^5$}
\end{array}\nonumber
\renewcommand{\arraystretch}{1.0}
\end{eqnarray}

\noindent
Hence non-zero values of the charges $P_i$ and $W_i$ in Matrix theory
indicate the presence of longitudinally wrapped M2 and M5 branes. 
Finite energy M-theory configurations are therefore selected by imposing
level matching constraints $P_i= 0$ and $W_i = 0$, which in the large $N$ 
NS 5-brane system (the last two columns)  project onto the sector with 
vanishing KK momenta and NS winding number.  The same level matching 
were introduced already in section 3 from the condition that the space-time 
supersymmetry algebra (\ref{susy}) be reproduced.

\medskip

\section{The Matrix M5-brane}

Let us now  put the description of sections 2 and 3
of  the M5-brane in a proper Matrix theoretic perspective.
The correspondence will be most clear in the formulation
of Matrix theory as the large $N$ limit of $N$ IIA 5-branes.

The fields parametrizing the low energy collective modes of a 
single IIA 5-brane form an ${\cal N} \! = \! (2,0)$ 
tensor multiplet identical to that used in sections 2 and 3. 
For the compactification on $\tilde{T}^5$, four of the five scalars 
represent the transverse oscillations of the
5-brane in non-compact directions, while one scalar field $Y$ is
extracted from one of the RR fields and is indeed periodic.
For $N$ IIA 5-branes, as long as they are far apart, we can separate
the worldvolume theory into $N$ different (2,0) tensor multiplets.
When they come close together, however, new types of
interactions occur. A precise low energy description of this 
interacting  theory is unfortunately not yet available at this point, 
although some new insights into its structure have been obtained.\footnote{
An (incomplete) list of references include  \cite{paulandy}\cite{sdstring}
\cite{mbh} \cite{nati} \cite{abkss}\cite{withiggs}.}

Particularly useful information can be extracted by considering
the IIA NS 5-branes from the perspective of M-theory\footnote{Here
in fact we do {\it not} mean the M(atrix) theory perspective related
via the $T_5 \; S \; T_5$ duality chain of section
3, but rather the M-theory whose $S_1$ compactification directly
gives the IIA theory of the last column of Table 1 in section 4.}. 
As first pointed out in \cite{paulandy},
M2-branes can end on M5-branes via open boundaries in the form
of closed strings attached to the M5-brane worldvolume.
These membrane boundaries are charged with respect to the
self-dual 3-form field strength $T=dU$ on the worldbrane.
If we consider an array of $N$ M5-branes, we thus deduce
that there are (at least) two basic types of string-like objects
on the 5+1-dimensional world volume: strings with variable 
tension that come from membranes stretched between two different 5-branes, 
and secondly, strings with fixed tension that correspond to membranes 
stretched around the 11th direction and returning to the same 5-brane.
The first type of strings are charged with respect to (the
difference of) the 3-forms of the two 5-branes that are connected
by the 2-brane. These charged strings are often referred to as
tension-less or self-dual strings. The second type of
strings, on the other hand, are neutral with respect to the
self-dual 3-form. They represent the micro-strings, \ie
the fractionated fundamental IIA strings trapped inside the 
worldvolume of the $N$ NS fivebranes. The collection of all
degrees of freedom together look like an (as yet unknown) 
non-abelian generalization of the abelian (2,0) theory. 

Charge conservation
tells us that the charged strings with non-trivial winding 
numbers around the $T^5$ are stable against breaking
and unwinding. For the neutral strings, on the other hand, 
there appears to be no similar world-brane argument 
available that would demonstrate their dynamical stability.
From the space-time perspective, however, the micro-string
winding number is identified with the total NS string
winding number (times $N$), and therefore represents a 
genuine conserved charge. This conserved quantity is the 
vector central charge $W_i$ discussed above.

Among the low energy degrees of freedom of the $N$ IIA 5-brane
bound state, we can isolate an overall
(2,0) tensor multiplet, that describes the collective center of mass
motion of the 5-branes. With the help of this decoupled tensor 
multiplet, we can include the configurations with non-vanishing 
D-brane charges by turning on the three types of fluxes $m_{ij}$, 
$p_i$ and $\qf$ as described in section 2. The respective IIA and 
 M-theory interpretation of these fluxes are as indicated in table 1 
in section 4. In this way we see that Matrix theory indeed exactly 
reproduces the description in section 2. In addition, we find that
the micro-string degrees of freedom, that were first postulated on the 
basis of U-duality symmetry, have now found a concrete origin.

Based on this new Matrix perspective, we can now obtain a much 
more detailed description of the M5-brane theory, based on the following 
proposed Matrix definition of its world-volume theory:\\

\noindent
{\it The world-volume theory of the M5-brane at light-cone 
momentum $p_+ = N/R$ is identical to that of a bound state of $N$ IIA 
NS 5-branes, subject to ten level matching constraints (\ref{levelmatch}). 
The KK momenta, M2 and M5-brane wrapping numbers are in this correspondence
identified with respectively the D4, D2 and D0-brane charges.}

\medskip

\noindent
In essence this identification amounts to a definition of the M5-brane as the
collection of all pointlike bound states of Matrix M-theory on
$\widetilde{T}^5$, including those with zero fivebrane wrapping number. 
Via this proposed definition, all the properties postulated in 
\cite{fivebrane} have become manifest.  The physical motivation 
and interpretation of the above proposal has been discussed already 
in section 2.  It would clearly be worthwhile to investigate
the correspondence between this characterization of the M5-brane 
world volume theory and the more conventional 
low-energy formulation obtained from 11-dimensional 
supergravity \cite{wittenfb}.

In the following sections we will to obtain some more
detailed dynamical information about the
5+1-dimensional micro-string theory.

\medskip

\section{$S_N$ Orbifolds and 2nd Quantized Strings}

Before continuing our journey through the matrix world of M-theory, 
it will be useful to present a mathematical identity that will be 
instrumental in establishing a precise link between the present formalism
for M-theory and the standard geometric framework of perturbative string 
theory.

Consider the supersymmetric sigma model on the orbifold target space,
defined by the symmetric product space
\be
S^NM = M^N/S_N.
\ee
Here $M$ denotes some target manifold, and $S_N$ denotes
the symmetric group, the permutation group of $N$ elements. 
The two-dimensional world-sheet is taken to be a cylinder parametrized 
by coordinates $(\sigma,\tau)$ with $\sigma$ between $0$ and $2\pi$. 
The correspondence we will show is that this sigma model, in the limit 
$N\ra \infty$, provides
an {\it exact} light-cone gauge 
description of the Hilbert space of {\it second quantized}
string theory on the target space-time $M \times \R^{1,1}$.

The $S_N$ symmetry is a gauge symmetry of the orbifold model, that
permutes all $N$ coordinate
fields $X_{I} \in M$, where $I=1, \ldots, N$ labeles the various copies
of $M$ in the product $M^N$.  Therefore we can define twisted sectors
in which, if we go around the space-like $S^1$ of the
world-sheet, the fields $X_I(\sigma)$ are multivalued, via
\be
X_I(\sigma + 2\pi) = X_{g(I)}(\sigma) 
\ee
where $g$ denotes an element of the  $S_N$ permutation group.  
A schematic depiction of these twisted sectors is
given in
\addtocounter{fignum}{1}
{\it fig.}\ \figuurnum\
and correspond to configurations with strings with various lengths.
\begin{figure}[t]\begin{center}
\leavevmode\hbox{\epsfxsize=7cm \epsffile{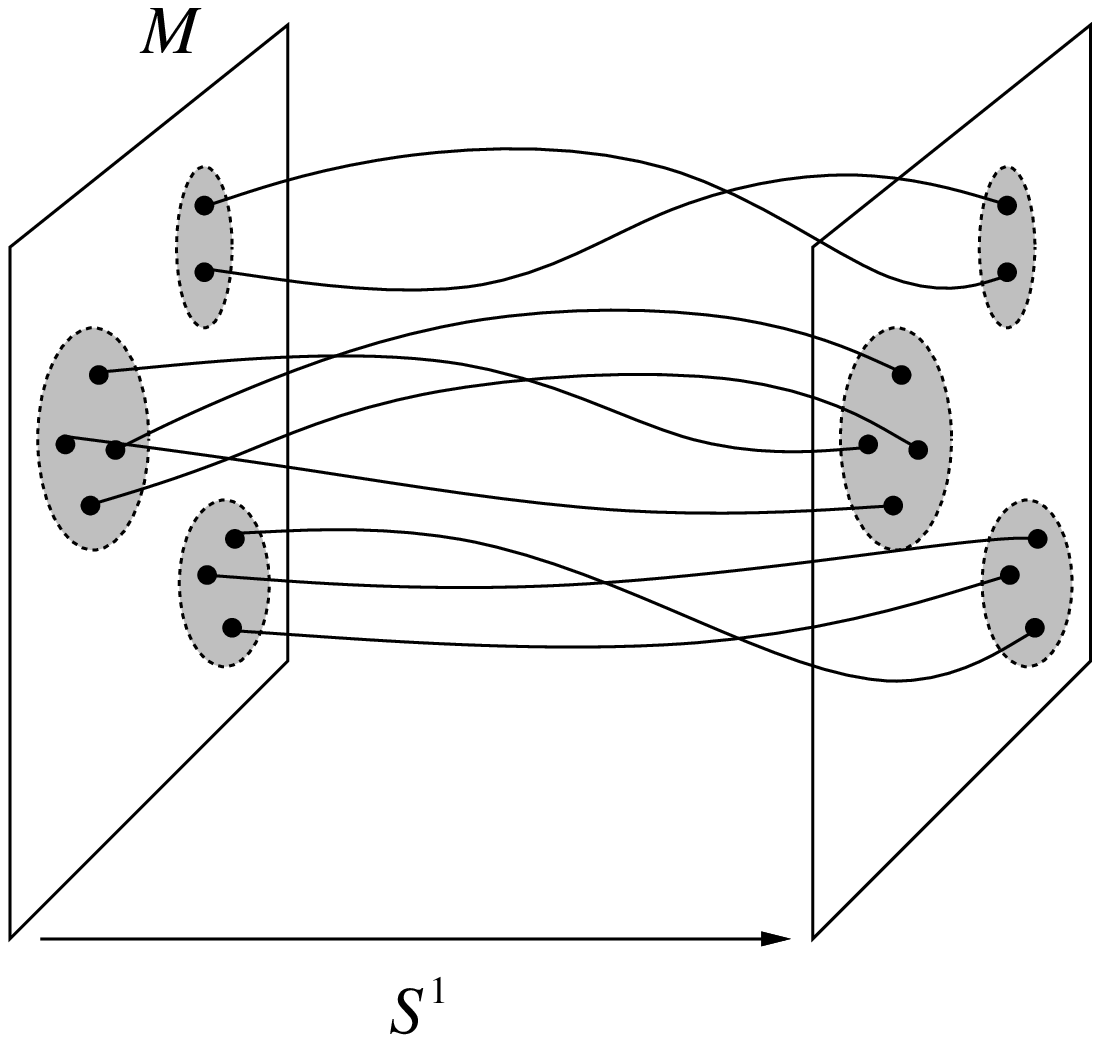}}\\
\parbox{7cm}{\small \bf Fig.\ \figuurnum: \it 
A twisted sector 
corresponds to a configuration with strings of various lengths.}
\vspace{-8mm}
\end{center} \end{figure}

The Hilbert space of this $S_N$ orbifold field theory is thus
decomposed into twisted sectors labeled by the conjugacy classes
of the orbifold group $S_N$  \cite{orbifold},
\be
\cH(S^NM) = \bigoplus_{{\rm partitions} \ \{N_n\}} \cH_{\{N_n\}}.
\ee
Here we used that for 
the symmetric group, the conjugacy classes $[g]$  are characterized
by partitions $\{N_n\}$ of $N$
\be
\sum_{n} n N_n = N,
\ee
where $N_n$ denotes the multiplicity of the cyclic permutation $(n)$
of $n$ elements in the decomposition of $g$
\be
[g] = (1)^{N_1}(2)^{N_2} \ldots (s)^{N_s}.
\ee
In each twisted sector, one must further keep only the
states invariant under the centralizer subgroup $C_g$ of $g$,
which takes the form
\be
C_g = 
\prod_{n=1}^s \, S_{N_n} \times \Z^{N_n}_n.
\ee
Here each factor $S_{N_n}$ permutes the $N_n$ cycles $(n)$, while each
$\Z_n$ acts within one particular cycle $(n)$.

Corresponding to this factorisation of $[g]$, we can decompose each
twisted sector into the product over the subfactors $(n)$ of
$N_n$-fold symmetric tensor products of appropriate smaller Hilbert
spaces $\cH_{(n)}$
\be
{\cH}_{\{N_n\}} = \bigotimes_{n>0} \, \Sym^{N_n} \cH_{(n)}
\ee
where\footnote{Here 
the symmetrization is assumed to be compatible with the grading
of $\cH$. In particular for pure odd states $S^N$
corresponds to the exterior product $\wedge^N$.}
\be
\Sym^N\cH = \Bigl(\underbrace{\cH \otimes \ldots \otimes \cH}_{N\ times}
\Bigr)^{S_N}.
\ee
The spaces $\cH_{(n)}$ 
in the above decomposition denote
the $\Z_n$ invariant subsector of the space of states of a
single string on $\R^8\times S^1$ with winding number $n$. We can
represent this space via a sigma model of $n$
coordinate fields $X_I(\sigma) \in M$ with the cyclic boundary
condition
\be
\label{boundaryc}
X_I(\sigma+2\pi) = X_{I+1}(\sigma), 
\ee
for $I \in (1, \ldots, n)$.
We can glue the $n$ coordinate fields $X(\sigma)$ together into one
single field $X(\sigma)$ defined on the interval $0\leq
\sigma \leq 2\pi n$. Hence, relative to the string with winding number
one, the oscillators of the long string that generate $\cH_{(n)}$ have a
fractional  ${1\over n}$ moding. 
The group $\Z_n$ is generated by the cyclic permutation
\be 
\omega : \ X_I \rightarrow X_{I+1}
\ee
which via (\ref{boundaryc}) corresponds to a translation $\sigma \ra 
\sigma + 2\pi$. Thus the $\Z_n$-invariant subspace consists of those states 
for which the fractional left-moving minus right-moving oscillator numbers 
combined add up to an integer. 

It is instructive to describe the implications of this structure
for the Virasoro generators $L_0^{(i)}$ of the individual strings.
The total $L_0$ operator of the $S_N$ orbifold CFT in a twisted
sector given by cyclic permutations of length $n_i$ decomposes
as
\be 
L^{tot}_0 = \sum_i {L_0^{(i)}\over n_i}.
\ee
Here $L_0^{(i)}$ is the usual canonically normalized operator in terms
of the single string coordinates $X(\sigma),\th(\sigma)$ defined above. 
The meaning of the above described $\Z_n$ projections is that it requires 
that the contribution from a single string sector to the total world-sheet
translation generator $L{}^{tot}_0-\Lbar{}^{tot}_0$ is integer-valued. 
From the individual string perspective, this means that 
$L{}^{(i)}_0-\Lbar{}^{(i)}_0$ is a multiple of $n_i$.

To recover the Fock space of the second-quantized type IIA string we
now consider the following large $N$ limit. We send $N\ra \infty$ and
consider twisted sectors that typically consist of a finite number of
cycles, with individual lengths $n_i$ that also tend to infinity in
proportion with $N$. The finite ratio $n_i/N$ then represents the
fraction of the total $p_+$ momentum (which we will normalize to 
$p_+^{tot}=1$) carried by the corresponding string\footnote{The fact that
the length $n_i$ of the individual strings specifies its light-cone
momentum is familiar from the usual light-cone formulation of 
string theory.}
\be
p_+^{(i)}= {n_i\over N}.
\ee
So, in the above terminology, only long strings survive. The
usual oscillation states of these strings are generated in the orbifold CFT 
by creation modes $\a_{-k/N}$ with $k$ finite. Therefore, in the large 
$N$ limit only the very low-energy IR excitations of the $S_N$-orbifold
CFT correspond to string states at finite mass levels. 

The $\Z_n$ projection discussed above in this limit effectively
amounts to the usual uncompactified level-matching conditions $L{}^{(i)}_0 -
\Lbar{}^{(i)}_0=0$ for the {\it individual} strings, since all single string
states for which $L_0-\overline{L}_0 \not=0$ become infinitely massive at
large $N$. The total mass-shell condition reads (here we put the string
tension equal to $\alpha' = 1$)
\be
p_-^{tot}=NL_0^{tot}
\ee
The mass-shell conditions of the individual strings is recovered 
by defining the individual light-cone momenta via
the usual relation 
\be
p_+^{(i)}p_-^{(i)}=L_0^{(i)}.
\ee 
All strings therefore indeed have the same string tension.

In the case that $M = \R^8$, we can choose the fields $X^i$, 
$\theta^a$, $\th^{\dot a}$ to transform respectively 
in the ${\bf 8}_v$ vector, and ${\bf 8}_s$ and ${\bf 8}_c$ spinor 
representations of the $SO(8)$ R-symmetry group of transversal rotations.  
The orbifold model then becomes exactly equivalent to the free string limit 
of the uncompactified type IIA string theory in the Green-Schwarz
light-cone formulation. The spacetime supercharges are given by 
(in the normalization $p_+^{tot} = 1$)
\ba
Q^a 
\is {1\over \sqrt{N}}\oint \!d\sigma\,  \sum_{I=1}^N \th_I^a, \nonu
Q^\adot \is \sqrt{N} \oint \!d\sigma\, \sum_{I=1}^N G_I^\adot, \nonumber
\ea
with
\be
G_I^\adot(z) =  \gamma^i_{a\adot} \th_I^a \d x_I^i
\ee
In the $S_N$ orbifold theory,  the twisted sectors of the orbifold 
corresponding to the possible multi-string states are all superselection 
sectors. We can introduce string interactions
by means of a local perturbation of the orbifold CFT, that generates 
the elementary joining and splitting of strings.

\medskip

\section{String Interactions via Twist Fields}

Consider the operation that connects 
two different sectors labeled by $S_N$ group elements that are related 
by multiplication by a simple transposition. It is relatively easy to 
see that this simple transposition connects a state with, say, one string 
represented by a cycle $(n)$ decays into a state with two strings 
represented by a permutation that is a product of two cycles $(n_1)(n_2)$ 
with $n_1+n_2=n$, or vice versa. So the numbers of incoming and outgoing 
strings differ by one. Pictorially, what takes place is that
the two coinciding coordinates, say  $X_1$ and $X_2 \in M$, 
connect or disconnect at the intersection point, and as illustrated in 
\figuurplus{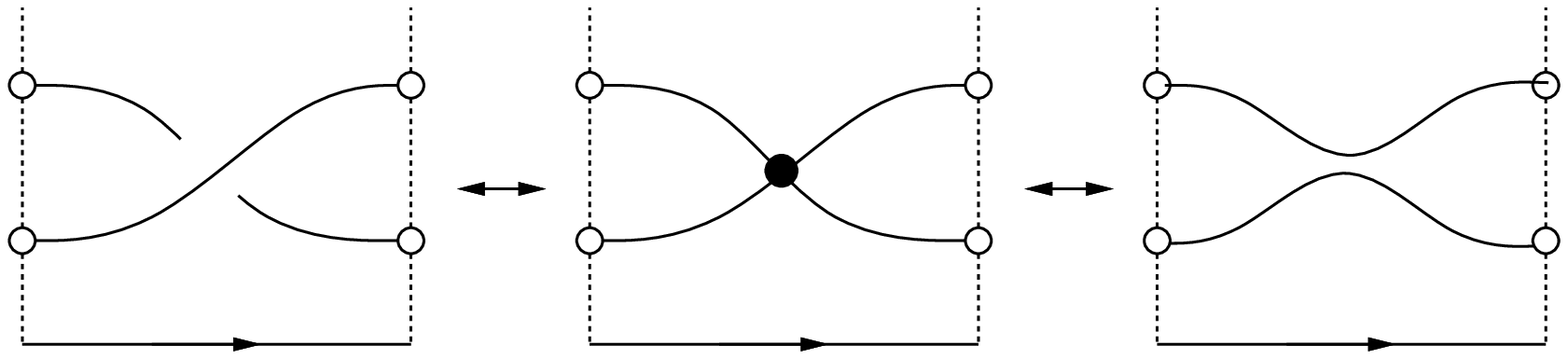}{7.6cm}{The splitting 
and joining of strings via a simple permutation.}
this indeed represents an elementary string interaction.

In the CFT this interaction is represented by a local operator, and thus
according to this physical picture one may view the interacting string 
theory as obtained via a perturbation of the $S_N$-orbifold conformal 
field theory. In first order, this perturbation is described via a 
modification of the CFT action 
\be
S = S_{CFT} + \lambda \int\! d^2z \; V_{int}
\ee
where $V_{int}$ is an appropriate twist operator, that generates the just 
described simple transposition of string coordinates.

Let us consider this operator in more detail for $M=\R^8$.
It is clear from the above discussion that the interaction vertex
$V_{int}$ will be a twist field that interchanges two eigenvalues, say
$X_1$ and $X_2$. It acts therefore as a $\Z_2$ reflection on the
relative coordinate $X_-=X_1-X_2$, which has 8 components. So we are 
essentially dealing with a standard $\R^8/\Z_2$ supersymmetric orbifold.
This orbifold has well-known twist fields, which
will receive contributions of the bosons and fermions.
The bosonic twist operator $\sigma$ is defined via the
operator product
\be
\partial X_-^i(z) \; \sigma(0) \! \sim\! z^{-{1\over 2}} \tau^i(0)
\ee
Since the fermions transform in spinor representation ${\bf 8}_s$, their
spin fields $\Sigma^i,\Sigma^\adot$ will transform in the vector
representation ${\bf 8}_v$ and the conjugated spinor representation
${\bf 8}_c$. They are related via the operator products
\be
\th_-^a(z) \; \Sigma^i(0)\! \sim\!  
z^{-{1\over 2}} \gamma^i_{a\adot}\Sigma^\adot(0)  
\ee
The bosonic twist field $\sigma$ and the spin fields 
$\Sigma^i$ or $\Sigma^\adot$ have all conformal dimension $h=\half$ (i.e.
${1\over 16}$ for each coordinate), and the conjugated field $\tau^i$ 
therefore has dimension $h=1$. 
For the interaction vertex we propose the following
$SO(8)$ invariant operator
\be
\tau^i\Sigma^i
\ee
This twist field has weight $3/2$ and acts within the Ramond 
sector. It can be written as the $G_{-{1\over 2}}$ 
descendent of the chiral primary field $\sigma\Sigma^\adot$,
and furthermore satisfies
\be
[G^\adot_{-{1\over 2}},\tau^i\Sigma^i]= \d_z(\sigma\Sigma^\adot).
\ee
To obtain the complete form of the effective world-sheet interaction
term we have to tensor the left-moving and right-moving twist fields
and to sum over the pairs of $I,J$ labeling the two possible
eigenvalues that can be permuted by the $\Z_2$ twist
\be
V_{int} =   \sum_{I<J} \left(\tau^i\Sigma^i \otimes 
\overline{\tau}^j\overline{\Sigma}^j\right)_{IJ}.
\label{vertex}
\ee
This is a weight $({3\over 2},{3\over 2})$ conformal field. The
corresponding coupling constant $\lambda$ has therefore total
dimension $-1$ and the interaction will scale linear in $g_s$
just as needed.  
The interaction preserves the world-sheet ${\cal N}=(8,8)$ supersymmetry, 
as well as the broken space-time supersymmetries $Q^a$.

It is interesting to compare the above twist field interaction with
the conventional formalism of light-cone string theory.
In comparison with the Riemann surfaces picture of interacting strings
\cite{mandelstam}, one should note that around the interaction point 
the usual string world-sheet is actually a double cover of the 
local coordinate $z$, while in the NSR language the above operator insertion
also represents a picture changing operation.\footnote{It should also
be noted that a similar formulation of light-cone gauge bosonic string
theory can be given in terms of an $S^N R^{24}$ orbifold. In this case
the twist field also carries conformal weight (3/2,3/2) \cite{rey}. In the recent
paper \cite{frolov} it was shown that this formulation indeed
exactly reproduces the Virasoro four-point amplitude. A very similar 
formulation of string perturbation theory was proposed among others
by V.Knizhnik \cite{knizhnik}.}

We expect that for more general transverse target spaces $M$ 
the interaction vertex can be constructed quite analogously.
Its conformal scale dimension depends on the dimension $d$
of $M$, or equivalently, the central charge of the corresponding 
${\cal N}\! =\! 1$ superconformal field theory, via
\be
\label{dscale}
h = {(d + 4)\over 8}.   
\ee
In particular we see that the interaction vertex is irrelevant only as
long as $d> 4$, and becomes marginal at $d=4$.

\section{Matrix String Theory}

Let us now return to the maximally supersymmetric $U(N)$ 
Yang-Mills theory (\ref{susym}). For generality, let us consider
this model in $d+2$-dimensions on a space-time manifold of the form
\be
M_\YM = T^d \times S^1 \times \R
\ee
where $\R$ represents time. Strictly speaking, this SYM theory is 
well-defined only in dimensions $d+2\leq 4$. In the following, 
however, we would like to consider the model in a particular IR limit 
in the $S^1$ direction, in which the $S^1$ is taken much larger than 
the transverse torus $T^d$. In addition, we wish to combine this limit
with the large $N$ limit, similar to the one discussed in the previous
sections. Indeed we would like to claim that the dynamics of the 
maximally supersymmetric YM theory in this limit exactly reduces to 
that of a (weakly) interacting type IIA string theory on the target 
space time
\be
M_S = T^d \times \R^{10-d}.
\ee

This particular IR limit effectively amounts to a dimensional
reduction, in which the dynamics of the $d+2$ dimensional SYM theory
becomes essentially two-dimensional. It is natural and convenient,
therefore, to cast the $d+2$ SYM model in the form of a
two-dimensional SYM theory. This can be achieved by replacing all
fields by infinite dimensional matrices that include the fourier modes
in the transversal directions.  In particular, the $d$ transverse
components $A^i$ of the gauge potential in the original SYM model thus
become $d$ additional two-dimensional scalar fields $X^i$ with
$i=1,\ldots, d$ that represent the infinite set of matrix elements in
a fourier basis of the corresponding transverse covariant
YM-derivatives. The infinite dimensional matrices $X^i$ defined
via this procedure satisfy the recursion relations \cite{wati}
\ba
X^i_{mn} \is X^i_{(m-1)(n-1)},\qquad \ i > d\nonu
X^i_{mn} \is X^i_{(m-1)(n-1)},\qquad \ i \leq d, \  m\neq n\nonu
X^i_{nn} \is 2 \pi L +X^i_{(n-1)(n-1)}\qquad \ i \leq d\nonumber
\ea
With this definition, the $d+2$ dimensional SYM action can be reduced 
to the following (quasi) two-dimensional form
\ba
\label{2d}
S_{2D} \is \int \! {\rm tr}\, \Bigl((D_{\a}X^I)^2 +
\psi\Gamma^\a D_\a \psi + \lambda^2
F_{\a\b}^2 \nonu 
& & \quad
+{1\over \lambda} \psi\Gamma^I[X^I,\psi]+\frac{1}{2\lambda^2}[X^I,X^J]^2 \Bigr)
\nonumber
\ea
The interpretation of this ${\cal N}=8$ SYM model as a matrix
string theory is based on the identification of the set of 
eigenvalues of the matrix coordinates $X^i$ with the coordinates of 
the fundamental type IIA strings. In addition, the coupling $\lambda$ 
is given in terms of the string coupling $g_s$ as 
\be
\lambda^2=\a'g_s^2.
\ee
The dependence on the string coupling constant
$\lambda$ can be absorbed in the area dependence of the two-dimensional
SYM model. In this way $\lambda$ scales inversely with world-sheet length.
The free string at $\lambda=0$ is recovered in the IR limit.  In this IR
limit, the two-dimensional gauge theory model is strongly coupled and
we expect a nontrivial conformal field theory to describe the IR fixed
point.

It turns out that we can find this CFT description via the following 
rather naive reasoning. We first notice that in the $g_s=0$ limit,
the last two potential terms in the 2-dimensional action (\ref{2d}) 
effectively turn into constraints, requiring the matrix fields $X$ and 
$\theta$ to commute. This means that we can write the matrix coordinates
$X^i$ in a simultaneously diaganolized form
\be
\label{diago}
X^i =  U \diag(x_1^i, \ldots, x^i_N) U^{-1}
\ee 
with $U\in U(N)$. Here $U$ and all eigenvalues $x^i_I$ can of course still
depend on the world-sheet coordinates. 

Once we restrict ourselves to Higgs field configurations of the form
(\ref{diago}), the two-dimensional gauge field $A_\alpha$ decouples.
To make this explicit, we can use the YM gauge invariance to put
$U=1$ everywhere, which in particular implies that the YM-current
$j_\YM = [X^i,\partial_\alpha X_i]$ vanishes. The only non-trivial
effect that remains is that via the Higgs mechanism, the charged components
of $A_\alpha$ acquire a mass for as long as the eigenvalues $x_i$ are distinct
and thus decouple in the infra-red limit. This leads to a description 
of the infra-red SYM model in terms
of $N$ Green-Schwarz light-cone coordinates formed by the eigenvalues 
$x^i_I,\th^a_I,\th^\adot_I$ with $I=1,\ldots,N$.

The $U(N)$ local gauge invariance implies that there is still 
an infinite set of discrete gauge transformations that act on 
the diagonalized configurations (\ref{diago}). Part of these discrete
gauge transformation act via translations on the eigenvalues $x^i_I$ that
represent the transversal components of the gauge potential,
turning them into periodic variables. In addition, we must 
divide out by the action of the Weyl group elements of U(N), that 
act on the eigenvalues $x^i_I$ via $S_N$ permutations. Thus, via
this semi-classical reasoning, we deduce that the IR conformal
field theory is the supersymmetric sigma model on the orbifold 
target space
\be
(T^d \times \R^{8-d})^N/S_N.
\ee
When we combine this with the results described in the previous 
two sections, we  deduce
that the super-Yang-Mills model in the strict IR limit gives the
same Hilbert space as the free light-cone quantization of 
second quantized type IIA string theory.

The conservation of string number and individual string momenta
will be violated if we turn on the interactions of the 1+1-dimensional
SYM theory. When we relax the strict IR limit, one
gradually needs to include configurations in which the non-abelian
symmetry gets restored in some small space-time region: if at
some point in the $(\sigma,\tau)$ plane two eigenvalues $x_I$ and
$x_J$ coincide, we enter a phase where an unbroken $U(2)$ symmetry is
restored. We should thus expect that for non-zero $\lambda$, there
will be a non-zero transition amplitude between states that are
related by a simple transposition of these two eigenvalues.

We now claim that this joining and splitting process
is indeed first order in the coupling constant $\lambda$ as defined in the
SYM Lagrangian. Instead of deriving this directly in the
strongly coupled SYM theory, we can analyze the effective operator that
produces such an interaction in the IR conformal field theory.
The identification $\lambda \sim g_s \sqrt{\a'}$ 
requires that this local interaction vertex $V_{int}$ must have
scale dimension 
\be
(h, \overline{h}) = ({3\over 2},{3\over 2})
\label{weights}
\ee 
under left- and right-scale transformations
on the two-dimensional world-sheet. The analysis of the previous
section confirms that this is indeed the case.

As a further confirmation of this physical picture, it is relevant
to point out that there is  no classical barrier against 
the occurrence of string interactions in the IR limit. For every
string world-sheet topology, there exists a solution to the SYM equation
of motion of the form (\ref{diago}) such that the eigenvalues $x^i_I$
form this string worldsheet via an appropriate branched covering
of the cylinder $S^1 \times \R$. (An explicit description of these
solutions is given in \cite{wynter}). 
The suppression of string interactions in the IR field theory 
therefore entirely arises due to the quantum fluctuations around
these classical field configurations. 
This suggests that there
should exist a close relation between the moduli space of classical  
configurations of the two-dimensional SYM model in the IR regime and
the universal moduli space of all two-dimensional Riemann surfaces.

{} From this Yang-Mills set-up, it is straightforward to derive the 
correct scaling dimension of the twist-fields (that create the branch cuts) 
by computing the one-loop fluctuation determinant around such a non-trivial 
classical field configuration. To start with, this computation only involves
the eigenvalue fields $x^i_I$, and thereby reduces to the standard
analysis of twist field operators in orbifold CFT used above. In the present 
context, however, the orbifold CFT is automatically provided with an ultra-violet 
cut-off of the order $\lambda$ (as defined in (\ref{2d})), since this is the place 
where the non-abelian SYM dynamics takes over. This explains why the 
coupling $\lambda$ in (\ref{2d}) also determines the strength of the twist field 
perturbation away from the orbifold CFT.

It would clearly be interesting to make via this set-up a direct quantitative 
comparison with the standard perturbative SYM calculations. In 
\cite{probe}\cite{banks} \cite{beckers} it was shown that the SYM amplitudes
at weak coupling reproduce the tree level graviton exchange amplitude of eleven
dimensional supergravity. Non-renormalization theorems of the SYM-model should 
ensure that these specific amplitudes receive no higher order or non-perturbative 
corrections, so that we can extrapolate these results the IR limit. Via the 
comparison with our IR description of the same amplitudes, one can thus uniquely fix 
the ratio between the orbifold perturbation parameter (\ie the string coupling) 
and the coupling constant $\lambda$ in the two-dimensional SYM lagrangian \ref{2d}.

The result (\ref{weights}) for the conformal dimension of the
interaction vertex $V_{int}$ of course crucially depends on the number
of Higgs scalar fields being equal to 8. The fact that for this case
$V_{int}$ is indeed irrelevant in the IR limit provides the
justification afterwards of the above naive derivation of the orbifold
sigma model as describing the IR phase of the $U(N)$ SYM
model. Another way of stating the same result is that the $\R^8/\Z_2$
cannot be resolved to give a smooth Calabi-Yau space.

It is indeed interesting to try to 
repeat the same study for SYM models with less supercharges.
Concretely, we could consider the dimensional reduction of 
the ${\cal N}\! = \! 1$ SYM model in 5+1-dimensions, which will
give us a matrix string theory with $d=4$ transverse Higgs 
fields. Via the above reasoning we then deduce that in the
strict IR limit in the $S^1$ direction, the string interactions
are not turned off, since the interaction vertex $V_{int}$
is now exactly marginal. We will return to this case in the next 
section.

Finally we remark that, for SYM theories with even less supersymmetries,
the naive reasoning applied in this section in fact breaks down, 
since in this case the interaction vertex $V_{int}$ becomes a 
relevant perturbation.  Hence in this case the correct IR dynamics 
is no longer accurately represented by a (small) deformation of the 
naive orbifold IR field theory. Instead one expects that the twist 
field $V_{int}$ obtains a non-zero expectation value, and 
the IR theory presumably takes the form of some Ising type QFT.

\medskip

\section{Microstring Interactions}

A relatively concrete model of the micro-string theory can be
obtained by applying the reasoning of \cite{nati} to the 5+1 
SYM formulation of Matrix theory. Namely, in this formulation
we can describe longitudinal NS 5-branes (that extend in the 
light-cone direction) by means of YM instanton configurations
\cite{ganor}. Here the 5-brane number simply coincides with the 
instanton number $k$.\footnote{This representation follows from
the table in section 7, where the longitudinal M5 branes (via 
their IIA identification with D4-branes) get mapped by
the $T^5$ duality to D1 strings, which manifests themselves 
as instantons in the large $N$ D5 SYM theory.}
Similar as in \cite{nati}, we can then consider
the world-volume theory of these $k$ longitudinal NS 5-branes
in the limit of zero string coupling. As also explained above, 
this corresponds to taking the limit where one of the 
compactification circles becomes very large. The theory thus
again reduces to the IR limit a two-dimensional matrix string 
theory of the form (\ref{2d}), but now with matrices $X^I$
that carry a non-zero topological charge \cite{ganor}\cite{branes}
\be
k = \tr X_{[i} X_j X_k X_{l]}
\ee
representing the instanton number on the transverse $T^4$.
It can be seen that in this case, the IR limit of the matrix 
string theory (\ref{2d}) no longer becomes a free field theory 
but instead reduces to a sigma model on the moduli space 
${\cal M}_{N,k}(\hat{T}^4)$ of $k$ $U(N)$ instantons on 
the four-torus $\hat{T}^4$. Since this instanton moduli 
space is hyperk\"ahler, the corresponding supersymmetric sigma 
model is indeed conformally invariant.

In \cite{mbh} it was proposed that this sigma-model could
be used as a concrete way of representing the interacting 
micro-string theory on the world-volume of the $k$ 5-branes. 
Crucial for this interpretation is the fact that the moduli 
space of instanton configurations on $T^4$ with rank $N$, 
instanton number $k$, and fluxes $m$ can be thought of 
as a (hyperk\"ahler) deformation of the symmetric product 
space $S^{n}T^4$ with $n=Nk +\half m\wedge m$. Hence via 
the results of the previous sections, the sigma model on 
${\cal M}_{N,k}(\hat{T}^4)$ is indeed expected to behave 
in many respects as a light-cone description of a 5+1-dimensional 
interacting string theory.

The interpretation of  this sigma model as describing the micro-string
dynamics on $k$ five-branes becomes a bit more manifest
via the so-called Mukai-Nahm duality transformation. This
transformation provides an isometry
\be
{\cal M}_{N,k}(\hat T^4) = {\cal M}_{k,N}(T^4)
\ee
between the moduli space of $k$ $U(N)$ instantons on $\hat{T}^4$
to the moduli space of $N$ instantons on a $U(k)$ Yang-Mills theory 
on the dual four-torus $T^4$. This dual $U(k)$ SYM theory can be 
thought of as the world-volume theory of $k$ fivebranes wrapped 
around $T^4$, and the $N$ instantons describe the micro-string degrees 
of freedom on this world-volume. This dual language is most suitable 
for discussing the decompactification limit of the M-theory torus, which leads 
to the description of $k$ uncompactified fivebranes as a non-linear sigma model on
${\cal M}_{k,N}(\R^4)$. 

The fact that there exists a superconformal deformation of
the orbifold sigma model on $S^n T^4$ is in direct accordance
with the results of the previous section, since it is reasonable 
to suspect that the marginal operator that represents this deformation 
can be identified with an appropriate $\Z_2$ twist operator that 
generates the splitting and joining interactions among the microstrings. 
As the string in this case has 4 transverse dimensions, the twist-field 
indeed has total dimension 2. Below we would like to make this
description more explicit.

To this end, it will be useful to consider the situation in Matrix
theory, which describes  the $k$ 5-branes in combination with 
a single IIA matrix string. In fact, to simplify the following
discussion, we will give this string the minimal $p_+$ momentum,
that is we will represent it simply by adding one row to the matrix
coordinates $X^i$, so that we are dealing with a $U(N+1)$ matrix
theory. Next we write
\be 
X_i \rightarrow \twomatrix{Z_i}{Y_i}{\overline{Y}_i}{x_i} 
\label{ZYx}
\ee 
Here the matrix entry $x_i$ represents the coordinate of a
propagating type IIA string (with infinitesimal light-cone momentum
$p_+$), and the $U(N)$ part $Z_i$ describes the $k$ 5-branes.
Similarly, we can write 
\be 
\theta_a \rightarrow
\twomatrix{\zeta_a}{\eta_a}{\overline{\eta}_a}{\theta_a} 
\ee 
The idea is now to fix $Z_i$ to be some given instanton configuration 
on $T^4$ plus a diagonal part $z_I {\bf 1}_{N\times N}$ that describes 
the location of the 5-brane in the uncompactified space, and to 
consider the propagation of the IIA string in its background. 

This set-up will be useful in fact for studying two separate
regimes, each of which are quite interesting \cite{withiggs} \cite{dias}. 
First there is
the micro-string regime, which corresponds to the case where
the extra string $x$ is captured inside the $k$ NS 5-branes.
In this case, the $Y$ variable need 
to be thought of as the extra degrees
of freedom that extend the moduli space of $k$ $U(N)$ instantons
to that of $k$ $U(N+1)$ instantons. Secondly,  we can also
consider the regime where the IIA string described by the $x$-coordinate
is far enough removed
from the NS 5-branes so that it propagates just freely
on the background geometry of the 5-brane solitons.
This outside regime corresponds to the Coulomb branch of the
matrix string gauge-theory, while the micro-string regime
represents the Higgs branch.

For both regimes, it will in fact be useful to isolate from the
$Y$ degrees of freedom in (\ref{ZYx}) the part that describes
the constant zero-modes in the instanton background,  satisfying 
\be
Z_{[i} Y_{j]} = 0, \qquad \quad \gamma^i_{\a\b} Z_i \eta^\beta = 0.
\ee
Since $Z_i$ corresponds to a $k$-instanton configuration we can use 
the index theorem to determine that there are $k$ such zero-modes,
both for the $Y$ and $\eta$ field. In the Coulomb phase they describe
the low energy degrees of freedom that mediate the interaction 
between the string and the 5-brane. In this phase, the $Y$-modes
are massive and thus are naturally integrated out in taking
the IR limit on the $S^1$. In the Higgs phase, on the other hand, 
they become massless two-dimensional fields, and  represent the
additional micro-string degrees of freedom that arise from the 
IIA string trapped inside the $k$ 5-branes. Notice that in this 
transition from the Coulomb phase to the Higgs phase the single IIA
string indeed ``fractionates'' into $k$ individual micro-strings.

The quadratic part of the action for the $Y$ degrees of freedom
takes the form
\ba
\label{linear}
S \is \int \! d^2\! \sigma \; \Bigl(\, (D {Y}_i)^2 +
|x_I\!-\! z_{I}|^2 | Y_{i}|^2\, \Bigr) \nonu
&& \  + \; \int d^2\!\sigma \; 
\Bigl( \, \overline{\eta}^\a_a D_+\eta^\a_a + 
\overline{\eta}^{\dot\a}_a D_-\eta^{\dot\a}_a  
\\[2.5mm] & & \qquad \qquad + \overline{\eta}_a^\a
\gamma_{\a\dot\b}^I (x\! -\! z)_I )\, \eta^{\dot\b}_a \; \Bigr) 
\nonumber
\ea
Here we suppressed the summation over the index labeling the 
$k$ zero-modes. The discrete symmetries of the instanton
moduli space imply that the model (\ref{linear}) possesses
a $S_k$ gauge symmetry that acts by permuting these $k$ coordinates.
In \cite{berkooz-douglas} the presence of the fivebrane background was
incorporated in matrix theory by adding an appropriate hypermultiplet
to the Matrix action. In our set-up, the role of these additional 
hypermultiplets is taken over by the off-diagonal fields $Y,\eta$.

Let us explain the fermion chiralities as they appear in the
Lagrangian (\ref{linear}).  
In the original matrix string action the $SO(8)$
space-time chiralities are directly correlated with the world-sheet
chirality, reflecting the type IIA string setup, giving left-moving
fermions $\theta^a$ and right-moving fermions $\theta^{\adot}$.  The
$T^4$ compactification breaks the $SO(8)$ into an internal $SO(4)$
that acts inside the four-torus times the transversal $SO(4)$
rotations in space-time. The spinors decompose correspondingly as
${\bf 8}_s \ra ({\bf 2}_+,{\bf 2}_+) \oplus ({\bf 2}_-,{\bf 2}_-)$ and
${\bf 8}_c \ra ({\bf 2}_+,{\bf 2}_-) \oplus ({\bf 2}_-,{\bf 2}_+)$.
The instanton on $T^4$ with positive charge $k>0$ allows only zero
modes of positive chirality, selecting the spinors that transform as
${\bf 2}_+$ of the first $SO(4)$ factor. So, through this process the
space-time chirality of the fermion zero modes gets correlated with
the world-sheet handedness, giving $k$ left-moving fields $\eta_a^\a$
and $k$ right-moving fields $\eta_a^{\dot\a}$, where $\a,\dot\a=1,2$
denote chiral space-time spinor indices and $a =1,2$ label chiral 
world-volume spinors.

The Higgs phase of the matrix string gauge theory corresponds to the regime
where $x_I-z_I \simeq 0$. The above lagrangian of the $4k$ 
string-coordinates $(Y,\eta)$ then represents a linearized description 
of the micro-string motion. inside the fivebrane worldvolume.
The linearized world-sheet theory has 4 left-moving and 
4 right-moving supercharges 
\be
\gG^{\dot{a}\a} =\oint
\del \xX^{\dot a b} \sS_b^\a, \qquad \ 
\overline{\gG}^{\dot a \dot\a}=\oint
\delbar \xX^{\dot a b} \overline{\sS}_b^{\dot\a}
\ee 
which correspond to the
unbroken part of the world-brane supersymmetry and satisfy
\be
\{ \gG^{\dot a\a} ,\gG^{\dot b\b}\} =
2\epsilon^{\dot a \dot b}
\epsilon^{\a \b} \, \lL_{0},
\ee
Here $\lL_{0}$ is the left-moving world-sheet Hamiltonian.

The ground states of the micro-strings
contain the low-energy collective modes of the five-brane.  
To make this explicit, we note that the left-moving ground state must form
a multiplet of the left-moving fermionic zero-mode algebra
\be
\{\sS_a^\alpha,\sS_b^\beta\}=\epsilon_{ab}\epsilon^{\alpha\beta}
\ee
which gives 2 left-moving bosonic ground states $|\alpha\rangle$ and 2
fermionic states $|a\rangle$. By taking the tensor product with the
right-moving vacua one obtains in total $16$ ground states, which
represent the massless tensor multiplet on the five-brane.
Specifically, we have states $|\a\rangle \otimes |\dot\beta\rangle$ 
that describe four scalars $X^{\alpha\dot\beta}=\sigma_i^{\alpha\dot\beta} X^i$
transforming a vector of the $SO(4)$ R-symmetry, while $|\alpha\rangle 
\otimes |a\rangle$ and $|a\rangle \otimes| {\dot\a}\rangle$ 
describe the fermions $\psi_a^\alpha$ and $\psi_a^{\dot\alpha}$.
Finally, the RR-like states $|ab\rangle$ decompose into the fifth scalar
$Y$, and the 3 helicity states of the self-dual tensor field.
These are indeed the fields that parametrize the collective
excitations of the NS-brane soliton\footnote{A similar CFT description 
of the fivebrane effective field theory has been proposed in \cite{kutasov}.}. 

This 5+1-dimensional string thus forms a rather direct
generalization of the 9+1-dimensional Green-Schwarz string. Clearly,
however, the linearized theory can not be the whole story, because
it would not result in a Lorentz covariant description of the M5 
world-brane dynamics. Micro string theory is indeed essentially 
an interacting theory, without a weak coupling limit. 
Its interacting dynamics is obtained by including the non-linear
corrections to (\ref{linear}), obtained by properly taking into
account the non-linear geometry of the moduli space  
${\cal M}_{N,k}(\widehat T^4)$ of $U(N)$ $k$-instantons on the 
four-torus $\hat T^4$ .

As argued above, the sigma model on ${\cal M}_{N,k}(\widehat{T}^4)$
should be representable in terms of a finite marginal deformation of 
the above model by means of a twist field. Concretely, the linearized 
orbifold model has four different twist fields
\be
V_{ab} = V_a \otimes \overline{V}_b
\ee
where the left-moving factor $V_a$ is 
\be
V_a = \gamma_{ab}^i \tau_i \Sigma^b
\ee
and a similar expression holds for the right-moving $\overline{V}_a$.
Here $\Sigma^b$ denotes the fermionic spin field, defined via the
operator product relation
\be
\sS^\a_a(z) \; \Sigma_b(0)
\! \sim\!
z^{-{1\over 2}} \epsilon_{ab} \tilde{\Sigma}^{\a}(0)
\ee
with $\tilde{\Sigma}^\a(0)$ the dual spin field. Both spin fields
have conformal dimension 1/4, while $\tau_i$ has dimension $3/4$.

So, all the above operators describe possible marginal deformations, 
that at least in first order conformal perturbation theory, preserve 
the conformal invariance of the sigma model QFT.
Theses 4 physical operators $V_{ab}$ are in one-to-one 
correspondence with the RR-like micro-string ground states 
$|a\rangle \otimes |b\rangle$, 
that represent the excitations of $T=dU$ and the
 scalar field $Y$. Rotation symmetry on the world-volume, however,
selects out the scalar combination 
\be
V_Y = \epsilon^{ab} V_a \otimes \overline{V}_b
\ee
that describes the emission of a $Y$ mode, as the only good 
candidate for representing the string interactions. 

Geometrically, the twist fields $V_{ab}$ represent the four
independent blow-up modes of a $\R^4/\Z_2$ orbifold singularity. The
three symmetric combinations deform the orbifold metric to a smooth
ALE space and the three parameters correspond to the different complex
structures that one can pick to specify the blow-up. These
deformations lead to regular supersymmetric sigma models, but are not
rotational invariant.  The antisymmetric combination $V_Y$ can be
interpreted as turning on a $B$-field $\theta$-angle dual to the
vanishing two-cycle around the $\Z_2$ orbifold point. h
The value of this $\theta$-angle away from the pure orbifold CFT,
which determines the strength of the micro-string interactions, 
is in principle fixed by the description of the matrix micro-string 
as the sigma-model on ${\cal M}_{N,k}(\hat T^4)$.

The value of the $\theta$-angle modulus
can be determined by considering the instanton moduli space around
the locus of point-like instantons. This problem was considered in
\cite{withiggs}, where it is proposed  that this naturally fixes its value
at $\theta=\pi$.\footnote{Here we defined $\theta=0$ as the value at the 
free orbifold CFT.} A puzzling aspect of this proposal is that this 
special value of $\theta$ is actually known to lead to a singular CFT 
\cite{aspinwall}. 
This is problematic, since it would make the sigma model description of the  
micro-string dynamics inconsistent, or at least incomplete. There could
however be a good physical reason for the occurrence of this singular CFT.
In the linear sigma model description of the Higgs branch, the singularity 
seems closely related to the presence of a Coulomb branch on the moduli 
space that describes the emission of strings from the fivebrane 
\cite{withiggs}. An alternative physical interpretation, however, is 
that the singularity of the 
CFT reflects the possible presence of the tensionless charged strings.

The latter interpretation in fact suggests that the singularity
may possibly be resolved by introducing a finite separation 
between the $k$ fivebranes. In the M-theory decompactification 
limit in particular, we know that the relevant moduli space 
${\cal M}_{k,N}(\R^4)$ allows for a number of deformation 
parameters in the form of the Higgs expectation values 
of the $U(k)$ SYM model. These Higgs parameters modify the
geometry of the instanton moduli space, and indeed have the
effect of softening most of its singularities. This is in direct 
accordance with the micro-string perspective, since for finite
Higgs expectation values the charged self-dual strings always 
have a finite tension.

\section{Future Directions}

To end with, we would like to mention some future directions,
where useful new insights may be obtained.

\medskip

\noindent
{\em Space-time Geometry.}

We have seen that the matrix setup is quite well adapted to extract
all the effective degrees of freedom that describe the type IIA
fivebrane, and one might hope that the present context of matrix
string theory opens up new ways of examining the interaction with
their environment.  In particular, at weak coupling one would like to
be able to recover the classical fivebrane geometry including its
anti-symmetric tensor field. This geometry should arise in the CFT
description of the Coulomb phase of the matrix string theory, where
the string $x(\sigma)$ propagates at a finite distance $|x_I- z_I|>0$
from the fivebrane soliton.

This description of the 
string and fivebrane is closely related to the recently studied 
case of a D-string probe in the
background of a D-fivebrane \cite{dstring-probe}.  
One can  derive an effective action for the $x$-part of the 
$U(N)$ matrix by integrating out the off-diagonal
components $Y$ and $\eta$. In \cite{dstring-probe} this 
was shown that the one-loop result for the 
transversal background metric
\be
ds^2 = \Bigl(\, 1+ {k\over |x-z|^2} \,\Bigr) (dx)^2. 
\ee
is indeed the well-known geometry of the NS fivebrane
\cite{callanetal}.

Most characteristic of the solitonic fivebrane is its magnetic charge 
with respect to the NS anti-symmetric tensor field $B$
\be
\label{H3}
\int_{S_3} H = k 
\ee 
where $H=dB$ for a three sphere enclosing the brane. This result 
can be recovered analogously, via a direct adaptation of the 
analysis of \cite{berkooz-douglas}. The transversal rotation 
group $SO(4)\simeq SU(2)_L\times SU(2)_R$ acts on the relative 
coordinate field $(x-z)^I$ via left and right multiplication.
This group action can be used to 
decompose the $2\times 2$ matrix $\gamma^{\a\dot\b}_I(x-z)^I$ 
in terms of a radial scalar field $r= e^\varphi$ and
group variables $g_L,g_R$ as 
\be
\gamma_I (x-z)^I = e^\varphi g_L g_R^{-1}.
\ee
The combination $g=g_Lg_R^{-1}$ labels the
$SO(3)$ angular coordinate in the transversal 4-plane.
Inserting the above expression for $x^I$ into (\ref{linear}),
the angles $g_L$ and $g_R$ can be absorbed into 
$\eta$ via a chiral rotation, producing a model of $k$ 
$SU(2)$ fermions coupled to
a gauge field $A_+=g_L^{-1}\d_+ g_L$, $A_-= g_R^{-1}\d_-
g_R$.  Via the standard chiral anomaly argument, one derives that 
the one-loop effective action includes an $SU(2)$ WZW-model for the 
field $g$ with level given by the five-brane number $k$. 
The background three-form field $H$ in (\ref{H3}) is reproduced as 
the Wess-Zumino term of this action. 

We thus recover the CFT description of the type IIA string moving
near the fivebrane \cite{callanetal}.  This action consists of a
level $k$ $SU(2)$ WZW model combined with a Liouville scalar field
$\varphi$
\be
S(x^I) = S_{WZW}(g) + S_{L}(\varphi). 
\ee
with
\be
\label{liou}
S_L(\varphi) = \int \! d^2 \! \sigma 
\Bigl( |\partial \varphi|^2 + \gamma R^{(2)} \varphi\, \Bigr)
\ee
The background charge $\gamma$ of the scalar $\varphi$ represents a 
space-time dilaton field that grows linearly for $\phi \ra - \infty$, 
and is normalized such that the total central charge is $c=6$.

It is quite meaningful  that the radial coordinate 
$r=e^{\varphi}$ is described in terms of a Liouville type
field theory (\ref{liou}). The role of $r$ is indeed very
similar to that of the world-sheet metric in two-dimensional
quantum gravity: since $r$ represents the mass of the lightest fields
that have been integrated out, it also parametrizes the 
short-distance world-sheet-length scale at which the CFT description 
is expected to break down. Besides the fact that it acquires a 
corresponding conformal scale dimension, this also means that 
it provides the natural cut-off scale at which the short-distance 
singularities need to be regulated.

In this light, it would be very interesting to study in more
detail the transition region between the Coulomb and Higgs phase
of this two-dimensional gauge theory, describing the capturing process
of the micro-string inside the fivebrane. The relevant
space-time geometry in this case should contain all the essential 
features of a black hole horizon, which initially is located at 
$r= e^\varphi = 0$. We expect that via the above CFT 
correspondence,  and in particular by studying the limits on its 
validity, useful new insights into the reliability of semi-classical 
studies of black hole physics can be obtained \cite{martinec-li,mbh}.

\medskip

\newcommand{\UU}{\! \mbox{\footnotesize\it  U}}%
\newcommand{\TT}{\! \mbox{\footnotesize\it  T}}%
\newcommand{\VV}{\! \mbox{\footnotesize\it  V}}%
\newcommand{\cF}{{\cal F}}%

\noindent
{\em Micro-strings on $K3 \times T^2$.}

Interesting new lessons can be learned from the beautiful 
geometric realization of heterotic string theory via M-theory 
compactified on $K3$. In this correspondence the perturbative 
heterotic string becomes a natural part of the M5-brane spectrum 
wrapped around the $K3$ \cite{harveystrominger}, 
which thereby provides a far reaching
generalisation of its world-sheet dynamics in terms of the above
5+1-dimensional micro-string theory. A matrix formulation
of the interactions between different wrapped M5-branes may 
ultimately provide a description of heterotic string dynamics,
in which non-perturbative duality is realized as a geometric
symmetry.

As a concrete illustration, we consider the 
generalization of the one-loop heterotic string partition function. 
In perturbative string theory, this one-loop calculation proceeds 
by considering the CFT free energy on a world-sheet with the
topology of $T^2$, \ie the closed world-trajectory of a closed 
string. In the M-theory realization, the string world-sheet
expands to become the world-volume of the M5-brane, while the
CFT get replaced by the micro-string theory. The one-loop 
diagram thus takes the form of a M5-brane world-trajectory
with the topology of $K3\times T^2$. The direct analogue 
of the perturbative string calculation thus involves the 
partition function second quantized micro-string theory
on this 6-dimensional manifold.

For the restricted case where one considers only the BPS 
contributions, this calculation can be explicitly done. 
The BPS restriction essentially amounts to a projection 
on only the left-moving micro-string modes, and thereby
eliminates the effect of its interactions. The M5-brane 
BPS partition sum can thus be computed by taking the 
exponent of the relevant one-loop free energy of the micro-string
\be
\label{part}
{\cal  Z}_{M5-brane} = \exp{{\cal F}^{micro-string}_{1-loop}} 
\ee
The above one-loop string partition function has in fact been 
computed already in \cite{kawai}. Let us briefly describe the result, 
because it in fact desplays some rather suggestive 
symmetries.

This partition sum ${\cal Z}$ will depend on the moduli
that parametrize the shape and size of world-volume manifold
$K3 \times T^2$. It turns out that there are three relevant
moduli $\TT$, $\UU$ and $\VV$, that parametrize the shape and size
of the $T^2$ and a Wilson line expectation value.
The expression for the partition function (\ref{part})
now takes the following form\footnote{
Here $(k,l,m)> 0$ means that $k, l\geq 0$ and $m\in \Z$, $m<0$ for
$k=l=0$.}
\ba
\label{product}
{\cal Z}(\TT,\UU,\VV)\! \! \is 
e^{-2\pi i(\TT +\UU +\VV)} \times
\\[2mm]
& & 
\!\!\!\!\!\!\!\!\!\!\!\!\!\!\!\!\!\!\!\!
\prod_{(k,l,m)> 0}\! \! \left(1 -e^{2\pi
i(k\TT+l\UU+m\VV)}\right)^{-c(4kl-m^2)}
\nonumber
\ea
where the coefficients $c(n)$ are defined by the expansion 
of the $K3$ elliptic genus as
\be
\chi_{\tau,z}(K3) = \sum_{h\geq 0,\, m\in \Z}
c(4h-m^2)e^{2\pi i (h\tau+mz)}
\label{K3-ell}
\ee
and are non-zero for $n\geq -1$.
The expression (\ref{product}) can indeed be recognized as a
second quantized string partition function \cite{orbifold}.
If we expand ${\cal Z}$ as a formal Fourier series expansion
\be
\label{expp}
{\cal Z}(\TT,\UU,\VV) = \sum_{k,l,m} D(k,l,m)
e^{-2\pi i (k\TT + l \UU + m \VV)}
\ee
the integer coefficients $D(k,l,m)$ represent the total 
(bosonic minus fermionic) number of states with given total 
momentum, winding and $U(1)$ charge, as specified by the three
integers $k,l,m$.

The above partition function can indeed be thought as the direct
generalization of the genus one partition function of perturbative
heterotic string BPS-states. The fact that ${\cal Z}$ now depends
on three moduli instead of one reflects the fact that the worldsheet
CFT is now replace by a world-volume micro-string theory, which in
particular also has winding sectors that are sensitive to the
size of the $T^2$. As a consequence, the partition function
${\cal Z}$ has the intriguing property that it is symmetric under
duality transformations, that in particular interchange the shape
and size parameters $\TT$ and $\VV$. 

The general group of dualities
is best exhibited by writing the three moduli into a 2 by 2 matrix
\be
\label{omega}
\Omega = \twomatrixd \TT \VV \VV \UU
\ee
which takes the suggestive form of a period matrix of a genus 2 curve.
The above partition function ${\cal Z}$ indeed
turns out to be a modular form under the corresponding $SO(3,2;\Z)$ 
modular group, and is in fact {\it invariant} under a $SL(2,\Z)$ 
subgroup. It is quite tempting to identify this $SL(2,\Z)$ symmetry 
with the electro-magnetic duality symmetry that arises upon further 
compactification of this theory to four dimensions. A concrete
suggestion that was made in \cite{dyon} is that the coefficients 
in the expansion (\ref{expp}) represent the degeneracy of 
(bosonic minus fermionic) dyonic BPS states with a given electric 
charge $q_e \in \Gamma^{22,4}$ and magnetic charge $q_m \in \Gamma^{22,4}$, 
via the identification
\be
d(q_m,q_e)=D(\half q_m^2, \half q_e^2,q_e\!\cdot\! q_m)
\ee
In \cite{dyon} it was shown that this formula survives a number
of non-trivial tests, such as correspondence with the perturbative
heterotic string as well as with the Bekenstein-Hawking black hole
entropy formula. If true, this result suggests that the 
presented formulation of M-theory via the microstring theory and
the M5-brane may in fact be extended to compactifications all the way 
down to four dimensions. Clearly, further study of matrix theory 
compactifications is needed to confirm or disprove this conjecture.

\bigskip

{\noindent  \sc Acknowledgements}

In the course of this work we benefited from useful discussions with
L. Alvarez-Gaum\'e, V. Balasubramanian, T. Banks, M. Becker,
C. Callan, M. Douglas, E. Kiritsis, I. Klebanov, D. Kutasov,
F. Larsen, J. Maldacena, L. Motl, E. Martinec, G. Moore, N. Seiberg,
S. Shenker, E. Silverstein,  A. Strominger, L. Susskind, W. Taylor,
P. Townsend, E. Witten, and others.  This research is partly supported
by a Pionier Fellowship of NWO, a Fellowship of the Royal Dutch
Academy of Sciences (K.N.A.W.), the Packard Foundation and the
A.P. Sloan Foundation.

\renewcommand{\Large}{\large}


\begin{thebibliography}{10}

\def\href#1#2{#2}
\bibitem{democracy}
P.~Townsend, ``P-brane democracy,'' in {\sl Particles, Strings
and Cosmology}, eds. J. Bagger e.a. (World Scientific 1996),  
hep-th/9507048.

\bibitem{witten}
E.~Witten, ``String {T}heory {D}ynamics in {V}arious {D}imensions,'' 
Nucl.  Phys. {\bf B443} (1995) 85--126,
\href{http://xxx.lanl.gov/abs/hep-th/9503124}{{\tt hep-th/9503124}}.

\bibitem{polch} J.~Polchinski, ``Dirichlet Branes and Ramond Ramond
Charges,'' {\tt hep-th/951001.}

\bibitem{fivebrane} R.~Dijkgraaf, E.~Verlinde and H.~Verlinde, ``BPS
Spectrum of the Five-Brane and Black Hole Entropy,'' Nucl. Phys.  {\bf
B486} (1997) 77--88, {\tt hep-th/9603126}; ``BPS
Quantization of the Five-Brane,'' Nucl. Phys. {\bf B486} (1997)
89--113, {\tt hep-th/9604055.}

\bibitem{bound} E.~Witten, ``Bound States of Strings and $p$-Branes,''
{\em Nucl. Phys.}  {\bf B460} (1996) 335--350,
\href{http://xxx.lanl.gov/abs/hep-th/9510135}{{\tt hep-th/9510135}}.

\bibitem{ulf} U.H.~Danielsson, G.~Ferretti, and B.~Sundborg,
``D-particle Dynamics and Bound States,'' {Int. J. Mod. Phys.}  {\bf
A11} (1996) 5463--5478,
\href{http://xxx.lanl.gov/abs/hep-th/9603081}{{\tt hep-th/9603081}};
D.~Kabat and P.~Pouliot, ``A Comment on Zerobrane Quantum Mechanics,''
Phys. Rev. Lett. {\bf 77} (1996) 1004--1007,
\href{http://xxx.lanl.gov/abs/hep-th/9603127}{{\tt hep-th/9603127}}.

\bibitem{probe}
M.~R. Douglas, D.~Kabat, P.~Pouliot, and S.~H. Shenker, ``D-branes and
Short Distances in String Theory,'' Nucl. Phys. {\bf B485} (1997)
85--127, \href{http://xxx.lanl.gov/abs/hep-th/9608024}{{\tt
hep-th/9608024}}.

\bibitem{hulltownsend} C. Hull and P. Townsend, ``Unity of 
Superstring Dualities,'' Nucl. Phys. {\bf B 438} (1995) 109.

\bibitem{banks} T.~Banks, W.~Fischler, S.~H. Shenker, and L.~Susskind,
``M Theory as a Matrix Model: A Conjecture,''
\href{http://xxx.lanl.gov/abs/hep-th/9610043}{{\tt hep-th/9610043}}.

\bibitem{motl}
L.~Motl, ``Proposals on Non-perturbative Superstring Interactions,''
{\tt hep-th/9701025}.
 
\bibitem{tomnati}
T.~Banks and N.~Seiberg, ``Strings from Matrices,'' {\tt hep-th/9702187}.

\bibitem{matrix-string} 
R. Dijkgraaf, E. Verlinde, and H. Verlinde, ``Matrix String Theory,'' 
{\tt hep-th/9703030.}

\bibitem{mbh}
R. Dijkgraaf, E. Verlinde, and H. Verlinde, ``5D Black Holes and
Matrix Strings,'' 
{\tt hep-th/9704018.}

\bibitem{nati} N.~Seiberg, ``Matrix Description of M-theory 
on $T^5$ and $T^5/Z_2$,'' {\tt hep-th/9705221}

\bibitem{hoppe}
J. Hoppe, ``Quantum Theory of a Relativistic Surface,'' Ph.D. Thesis,
MIT (1982).  B.\ de Wit, J.\ Hoppe and H.\ Nicolai, ``On the Quantum
Mechanics of Supermembranes,'' Nucl.\ Phys. {\bf B305} (1988)
545--581.



\bibitem{rozali}
M. Rozali, ``Matrix Theory and U-Duality in Seven Dimensions,''
{\tt hep-th/9702136}.

\bibitem{brs} 
M. Berkooz, M. Rozali, and N. Seiberg,
``Matrix Description of M theory on $T^4$ and $T^5$,''
{\tt hep-th/9704089}. 

\bibitem{shrink} W. Fishler, E. Halyo, A. Rajaraman, and L. Susskind, 
``The Incredible Shrinking Torus,'' {\tt hep-th/9703102.}

\bibitem{maldacena-susskind}
J.M. Maldacena and L. Susskind, ``D-branes and Fat Black Holes,''
Nucl. Phys. {\bf B475} (1996) 679, {\tt hep-th/9604042}.

\bibitem{branes}
T.~Banks, N.~Seiberg, and S.~Shenker, ``Branes from Matrices,''
\href{http://xxx.lanl.gov/abs/hep-th/9612157}{{\tt hep-th/9612157}}.

\bibitem{sb96} H. Verlinde, ``The Stringy Fivebrane,''  talk 
at Strings '96, Santa Barbara, June 1996.

\bibitem{stromingervafa} A. Strominger, C. Vafa, ``Microscopic Origin
of the Bekenstein-Hawking Entropy,''{{\tt hep-th/9601029.}}

\bibitem{juan} J. Maldacena, ``Statistical Entropy of Near
Extremal Five-branes,'' hep-th/9605016.

\bibitem{maldacena} For a recent account of these developments see
J. Maldacena and A. Strominger, ``Universal Low-Energy Dynamics for
Rotating Black Holes'' {\tt hep-th/9702015}, and references therein.

\bibitem{sdstring}
E. Witten, ``Some Comments on String Dynamics,'' hep-th/9510135;

\bibitem{paulandy} 
A. Strominger, ``Open P-Branes,'' {\tt hep-th/9512059};
P. Townsend, ``D-branes from M-branes,'' {\tt hep-th/9512062}.

\bibitem{candy} A. Losev, G. Moore, S. L. Shatashvili, ``M $\&$ m's,''
{\tt hep-th/9707250};
N. Seiberg and S. Sethi
``Comments on Neveu-Schwarz Five-Branes,''  {\tt hep-th/9708085} 

\bibitem{mfive}
P. Pasti, D. Sorokin, and M. Tonin,
``Covariant Action for a D=11 Five-Brane with the Chiral Field,''
Phys.Lett. {\bf B398} (1997) 41-46, {\tt hep-th/9701037};
M. Aganagic, J. Park, C. Popescu, J. Schwarz
``World-Volume Action of the M Theory Five-Brane,'' Nucl.Phys. 
{\bf B496} (1997) 191-214, {\tt hep-th/9701166};
J. Schwarz, ``Remarks on the M5-Brane,'' talk at 
Strings `97, {\tt hep-th/9707119}.

\bibitem{sorokintownsend}
D. Sorokin, P. Townsend, `` M-theory superalgebra from the M-5-brane,''
hep-th/9708003

\bibitem{wittenfb} E. Witten, ``Five-Brane Effective Action In M-Theory''
{\tt hep-th/9610234}

\bibitem{orbifold}
 R. Dijkgraaf, G, Moore, E. Verlinde, and H. Verlinde,
``Elliptic Genera of Symmetric Products and Second Quantized Strings,''
Commun. Math. Phys.  {\tt hep-th/9608096}.

\bibitem{mandelstam}
S.~Mandelstam, ``Lorentz Properties of the Three-String Vertex,''
Nucl. Phys. {\bf B83} (1974) 413--439; ``Interacting-String Picture of the 
Fermionic String,'' Prog. Theor. Phys. Suppl. {\bf 86} (1986) 163--170.

\bibitem{rey}
S-J. Rey, ``Heterotic M(atrix) Strings and Their Interactions,''
{\tt hep-th/9704158}.

\bibitem{frolov} G. Arutyunov and S. Frolov, 
``Virasoro amplitude from the $S^N\R^{24}$ orbifold sigma model,''
{\tt hep-th/9708129}.

\bibitem{wynter}
{T. Wynter, ``Gauge fields and interactions in matrix string theory,''
{\tt hep-th/9709029}}

\bibitem{beckers}
K. Becker and M. Becker, ``A Two Loop Test of M(atrix) Theory, {\tt hep-th/9705091}.\\
K. Becker, M. Becker, J. Polchinski, and A. Tseytlin, ``Higher Order Scattering
in M(atrix) Theory,'' {\tt hep-th/9706072}.

\bibitem{knizhnik}
V. Knizhnik, Sov. Phys. Usp. 32 (1989) 945.

\bibitem{wati}
W.~Taylor, ``D-brane Field Theory on Compact Spaces,''
\href{http://xxx.lanl.gov/abs/hep-th/9611042}{{\tt hep-th/9611042}}.

\bibitem{ganor}
O.J. Ganor, S. Ramgoolam, and W. Taylor, ``Branes, Fluxes and Duality in
M(atrix)-Theory,'' \href{http://xxx.lanl.gov/abs/hep-th/9611202}{{\tt
hep-th/9611202}}.

\bibitem{kutasov}
D. Kutasov, ``Orbifold and Solitons,'' {\tt hep-th/9512145}.

\bibitem{dstring-probe} M. Douglas, J. Polchinski, and A. Strominger, 
``Probing Five-Dimensional Black Holes with D-Branes,'' {\tt 
hep-th/9703031}.

\bibitem{berkooz-douglas}
M. Berkooz and M. Douglas, ``Fivebranes in M(atrix) Theory,''
{\tt hep-th/9610236.}

\bibitem{abkss}
O. Aharony, M. Berkooz, S. Kachru, N. Seiberg, and E. Silverstein, 
``Matrix Description of Interacting Theories in Six Dimensions,''
{\tt hep-th/9707079}

\bibitem{withiggs} E. Witten, ``On the Conformal Field Theory of the Higgs
Branch,'' {\tt hep-th/9707093} 
    
\bibitem{dias} D. Diaconescu and N. Seiberg,
 ``The Coulomb Branch of (4,4) Supersymmetric Field Theories in Two Dimensions,'' 
{\tt hep-th/9707158}

\bibitem{aspinwall} P. Aspinwal, ``Enhanced Gauge Symmetries and K3
Surfaces,'' Phys. Lett. {\bf B357} (1995) 329, {\tt hep-th/9507012}.

\bibitem{callanetal} C. Callan, J. Harvey and A. Strominger,
``World-Sheet Approach to Heterotic Solitons and Instantons,''
Nucl. Phys. {\bf 359} (1991) 611; 
I. Antoniadis, C. Bachas, J. Ellis and
D. Nanopoulos, ``Cosmological String Theories and Discrete
Inflation,'' Phys. Lett. {\bf 211B} (1988) 393; Nucl. Phys. {\bf B328}
(1989) 117; S.-J. Rey, ``Confining Phase of Superstrings and Axionic
Strings,'' Phys. Rev. {\bf D43} (1991) 439.

\bibitem{martinec-li}
M. Li and E. Martinec, ``Matrix Black Holes,'' {\tt hep-th/9703211};
talk at {\it Strings '97}, {\tt hep-th/9709114}.


\bibitem{harveystrominger} J. A. Harvey and A. Strominger, ``The
Heterotic String is a Soliton,'' Nucl. Phys. {\bf B 449} (1995)
535-552; Nucl. Phys. B458 (1996) 456-473, {\tt hep-th/9504047}.\\
S. Cherkis and J.H. Schwarz, ``Wrapping the M Theory Five-Brane
on K3,'' {\tt hep-th/9703062.}


\bibitem{dyon}
R. Dijkgraaf, E. Verlinde, and H. Verlinde, 
``Counting Dyons in N=4 String Theory,''
Nucl.Phys. {\bf B 484} (1997) 543, {\tt hep-th/9607026}
\bibitem{kawai} T. Kawai, ``$N=2$ Heterotic String Threshold
Correction, $K3$ Surface and Generalized Kac-Moody Superalgebra,''
hep-th/9512046.




\end{thebibliography}
\end{document}